\documentclass{aa}

\usepackage{psfig}
\def\void#1{{}}
\def\lsim{~\rlap{$<$}{\lower 1.0ex\hbox{$\sim$}}}
\def\gsim{~\rlap{$>$}{\lower 1.0ex\hbox{$\sim$}}}
\def\etal{{\it et al.\/}\ }
\def\ie{{\it i.e.\/}\rm,\ }
\def\eg{{\it e.g.\/}\rm,\ }

\usepackage{graphicx}

\usepackage{times}
\usepackage{mathptm}
\fontfamily{ptmcm}\selectfont
\usepackage{graphicx}
\usepackage{threeparttable}

\include{psfig.sty}

\begin{document}



\title{ESO Imaging Survey}{\subtitle{Deep Public Survey: Multi-Color
Optical  Data
for the Chandra Deep Field South\thanks{Based on observations
collected at the European Southern Observatory, La Silla, Chile within
program ESO 164.O-O561.}.}


\author{ S. Arnouts \inst{1}
 \and B. Vandame \inst{1} 
 \and C. Benoist \inst{2}
 \and M.A.T. Groenewegen \inst{1}
 \and L. da Costa \inst{1}
 \and M. Schirmer \inst{3}
 \and R. P. Mignani \inst{1}
 \and R. Slijkhuis \inst{1}
 \and E. Hatziminaoglou \inst{1}
 \and R. Hook \inst{1}
 \and R. Madejsky \inst{1,4}
 \and C. Rit\'e  \inst{5}
  \and A. Wicenec \inst{1}
}

\offprints{Stephane Arnouts, e-mail: sarnouts@eso.org}

\institute{
European Southern Observatory, Karl-Schwarzschild-Str. 2, D--85748
Garching b. M\"unchen, Germany
\and Observatoire de la C\^{o}te d'Azur, BP 229, 06304 Nice
cedex 4, France 
\and Max-Planck Institut f\"ur Astrophysik, Karl-Schwarzschild-Str. 1, D-85748 Garching b.  
     M\"unchen, Germany
\and  Universidade Estadual de Feira de Santana, Campus Universit\'ario,
            Feira de Santana, BA, Brazil
\and Observat\'orio Nacional, Rua Gen. Jos\'e Cristino 77, Rio de
Janerio, R.J., Brasil }

\date{received,  accepted}


\abstract  {This paper presents  multi-passband optical  data obtained
from observations of the Chandra Deep  Field South (CDF-S), located at
$\alpha   \sim 3^h   32^m$,  $\delta  \sim   -27^o  48^{\prime}$.  The
observations were conducted at the ESO/MPG  2.2m telescope at La~Silla
using  the 8k$\times$8k Wide-Field Imager (WFI).  This data set, taken
over a period of one year, represents the  first field to be completed
by the ongoing Deep Public Survey  (DPS) being carried  out by the ESO
Imaging Survey (EIS)   project.    This paper describes    the optical
observations,  the   techniques  employed  for  un-supervised pipeline
processing and   the  general  characteristics   of   the  final  data
set. Image processing has been  performed using multi-resolution image
decomposition techniques  adapted to the   EIS pipeline. The automatic
processing steps include  standard  de-bias and  flat-field, automatic
removal  of    satellite  tracks,  de-fringing/sky-subtraction,  image
stacking/mosaicking and  astrometry. Stacking  of dithered   images is
carried  out using pixel-based  astrometry which enables the efficient
removal of  cosmic rays and  image  defects, yielding remarkably clean
final   images. The  final  astrometric   calibration  is  based on  a
pre-release  of the GSC-II  catalog   and has an estimated   intrinsic
accuracy  of $\lsim0.10$~arcsec, with  all  passbands sharing the same
solution.   The paper  includes data  taken  in six  different filters
$U'UBVRI$. The   data  cover an  area   of about  0.25~square  degrees
reaching  5$\sigma$    limiting   magnitudes    of     $U'_{AB}=26.0$,
$U_{AB}=25.7$, $B_{AB}=26.4$, $V_{AB}=25.4, R_{AB}=25.5$ and $I_{AB}=$
24.7~mag, as measured within a  $2 \times$ FWHM aperture.  The optical
data covers the area of $\sim$ 0.1~square degrees for which moderately
deep  observations in  two  near-infrared  bands are  also  available,
reaching $5\sigma$ limiting    magnitudes of    $J_{AB}\sim23.4$   and
$K_{AB}\sim22.6$.  The current   optical/infrared   data  also   fully
encompass the region of the deep X-ray observations recently completed
by the Chandra telescope.  The optical data presented here, as well as
the infrared data  released earlier, are publicly available world-wide
in the form of  fully calibrated pixel and  associated weight maps and
source lists extracted in  each passband. These  data can be requested
through  the  URL  ``http://www.eso.org/eis''.  \keywords{catalogs  --
surveys-- stars: general - galaxies: general} } 

\maketitle

\section{Introduction} 
\label{sec:intro}

The successful application of photometric techniques to identify
high-redshift galaxies ($z \gsim 3$) combined with large-aperture
telescopes with multi-object spectrographs have provided the means to
study large-scale structures at high-redshifts and to directly probe
their evolution over a broad range of redshifts. Currently, the
challenge is to construct well-defined statistical galaxy samples at
faint magnitudes and use photometric redshift techniques to pre-select
candidates for large spectroscopic surveys probing the high-redshift
Universe. To achieve this goal, deep optical/infrared multi-passband
surveys covering areas of the order of a few degrees are required.

Interest in this type of dataset is widespread in Europe given the
imminent commissioning of the multi-object spectrograph VIMOS, part of
the new generation of VLT instruments. Foreseeing this need the
Working Group (WG) for public surveys at ESO recommended in early 1999
the undertaking of a deep, multi-passband optical survey, covering
three separate regions of sky one square degree each, with some areas
also covered in the infrared. These regions, referred to as Deep-1, 2
and 3, are located over a wide range of right ascension and each
correspond to four adjacent pointings (a-d) of the Wide-Field Imager
(WFI) mounted on the 2.2m MPG/ESO telescope at La~Silla. The main goal
of the survey is to produce suitable statistical samples to study
large-scale structures in the redshift domains $z\lsim1$ and
$z\gsim3$.  Details of the position of the selected regions can be
found at the URL ``http://www.eso.org/science/eis/''. One of the
selected regions includes the Chandra Deep Field South (CDF-S;
Giacconi \etal 2001) chosen because of its low column-density of
hydrogen and absence of bright stars, characteristics that make it
well-suited for deep X-ray observations with the Chandra (AXAF)
telescope. These observations have recently been completed for a total
integration time of one million seconds producing the deepest,
high-resolution X-ray image ever taken (Rosati \etal 2001).

In this paper the $UBVRI$ optical observations with the WFI of one of
the selected fields (Deep-2c), covering the CDF-S, are
presented. These data are used to make a first assessment of the
procedures and overall performance of the optical part of the Deep
Public Survey (DPS).  The optical data presented here complement those
in the near-infrared covering an area of $\sim0.11$~square~degrees
(Vandame \etal 2001) encompassing the region covered by the Chandra
observations. The new data reported here extend earlier
optical/infrared observations of this field carried out by EIS at the
NTT in 1998, covering a considerably smaller region. These older data
are currently being re-analyzed to produce, together with the new data,
a homogeneous dataset (Benoist \etal 2001a) which will supersede the
earlier 1998 release (Rengelink \etal
1998). Section~\ref{sec:observations} reviews the overall observing
strategy and describes the observations. Section~\ref{sec:reductions}
discusses the techniques employed in the data reduction and the
photometric and astrometric calibration of the images. While other
derived products are expected to be made available in the near future,
only photometric parameters of the sources detected in each of the
available passbands are presented for the moment in
Section~\ref{sec:catalogs}.  Even though the goal of the present paper
is not to interpret the data being released, Section~\ref{sec:results}
presents the results of comparisons between the present observations
and those of other authors. This is done for the sole purpose of
assessing the quality of the astrometry and photometry of the present
data set and its adequacy to meet the main goals of the
survey. Finally, a brief summary is presented in
Section~\ref{sec:summary}.

\section{Observations}
\label{sec:observations}

The optical observations of the CDF-S region ($\alpha \sim 3^h 32^m$,
$\delta \sim -27^o 48^{\prime}$) were carried out using the WFI camera
at the Cassegrain focus of the MPG/ES0 2.2m~telescope at the La~Silla
observatory. WFI is a focal reducer-type mosaic camera with $4 \times
2$ CCD chips. Each chip is composed of a $2048 \times 4098$ pixel
array and covers a field of view of $8.12 \times 16.25$ arcmin with a
projected pixel size of 0.238~arcsec. The 8 CCDs are physically
separated by gaps of width 23.8 and 14.3~arcsec along the right
ascension and declination directions, respectively. The full field of
view of the camera is thus $34\times 33$~arcmin, with a filling factor
of 95.9\%. Details on the WFI layout can be found, for instance, at
the URL
``http://www.eso.org/science/eis/eis\_proj/public/wfi\_chip.html'' as
well as at other locations on the ESO web pages.

Observations of the CDF-S have been performed using the WFI filters
summarized in Table~\ref{tab:filters}.  The $U-$band observations were
originally conducted using the available narrow filter which was
recently (October 2000) replaced by a broader filter to increase the
efficiency of the survey.  The transmission curves of these filters
are plotted in Figure~\ref{fig:transmission} together with their final
throughputs taking into account the optics and the CCD quantum
efficiency. It is worth mentioning that very little was known about
the performance of the instrument at the time the survey was
conceived, especially at the blue and red ends of the wavelength range
considered.

\begin{table}
\begin{center}
\caption{Filters used in the  observations of the CDF-S.} 
\begin{tabular}{llrr} \hline\hline
{\em Filter}   & {\em ESO id} & {\em $\lambda_{eff}$} &  {\em FWHM}   \\ 
               &          & (\AA)                 &  (\AA)          \\ \hline
U    & $U/38$     & 3710& 325    \\
U'   & $U 350/60$ & 3538   &  560   \\
B    & $B/99$     & 4556 & 905  \\
V    &  $V/89$    & 5342 & 900   \\
R    & $Rc/162$   & 6402  & 1595  \\
I    &  $Ic/lwp$  & 8535  & 1387  \\ \hline\hline 
\label{tab:filters}
\end{tabular}
\end{center}

\end{table}

\begin{figure} 
\centerline{\hbox{\psfig{figure=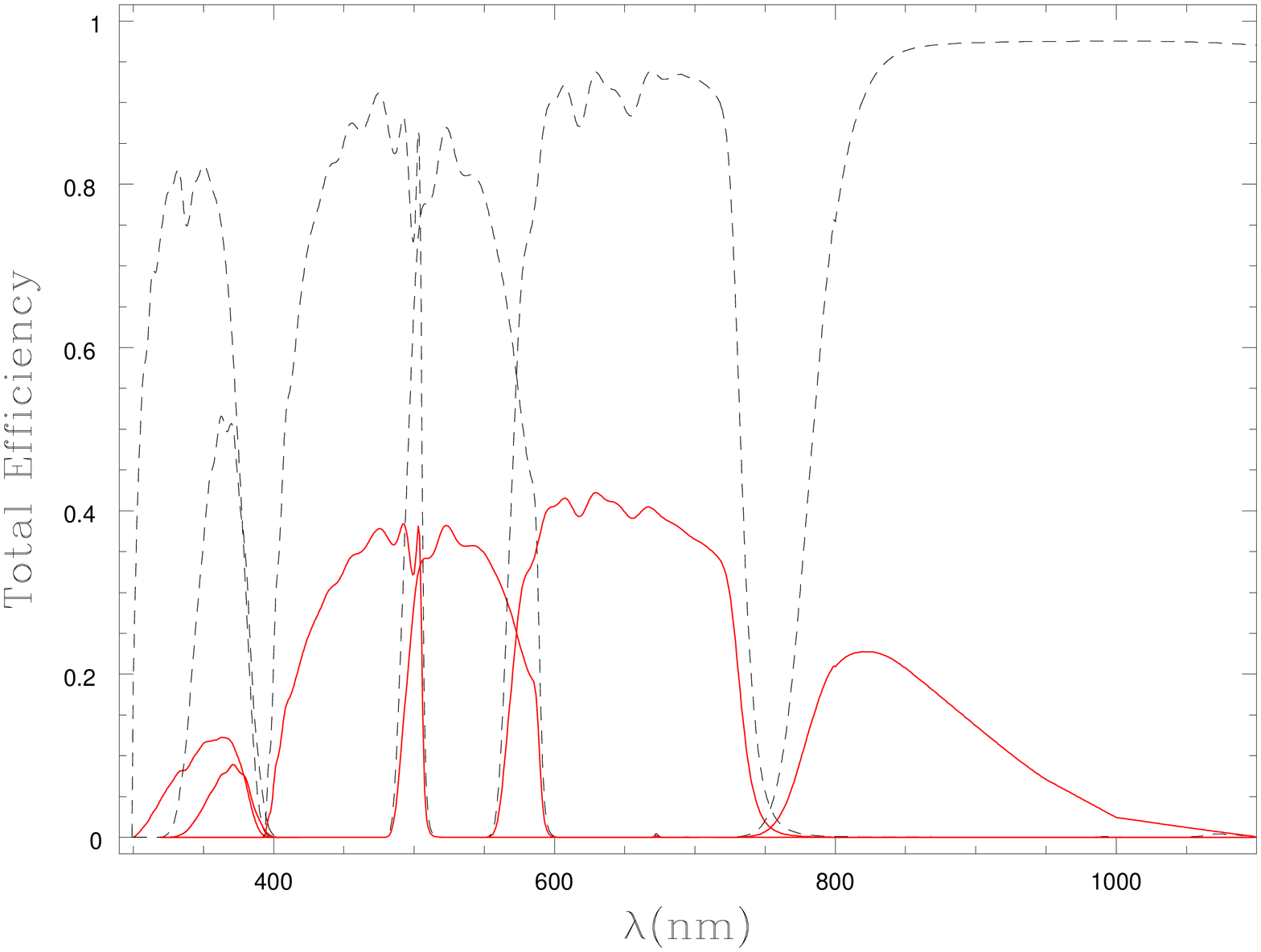,angle=-90,width=8.8cm,clip=}}}
\caption{Transmission curves of the filters (dashed line) and the total
throughput of the system (solid line).}
\label{fig:transmission}
\end{figure} 

The observations in each passband were split into observation blocks
(OBs) consisting of a sequence of five exposures with the integration
time for each individual exposure ranging from 5 to 15 minutes
depending on the passband. Table~\ref{tab:strategy} summarizes the
planned observational strategy listing: in column (1) the passband; in
column (2) the integration time of each individual exposure; in column
(3) the number of exposures per OB; in column (4) the integration time
per OB; in column (5) the number of OBs; and in column (6) the planned
total integration time per band. According to the planned strategy a
total of 210 WFI frames were required to complete a single pointing.
In order to minimize the imprint of the inter-chip gaps the choice of
pointings were done in the following way. First, the nominal position
of each OB (in a given passband) was drawn from a pre-defined list of
distinct positions, as given in Table~\ref{tab:mainpos}. For instance,
in the case of $U-$band which requires 13 OBs, the dithering pattern
used is illustrated in Figure~\ref{fig:pattern}.  Next, each exposure
in the OB sequence was dithered relative to the OB reference position
using the offsets in right ascension ($\Delta \alpha \cos \delta$) and
in declination ($\Delta \delta$) as listed in Table~\ref{tab:dither}
and illustrated in the inset of Figure~\ref{fig:pattern}.  This
dithering is essential for de-fringing as well as for efficiently
removing cosmic rays and other CCD blemishes.

\void{
\begin{table}[h]
\begin{center}
\caption{Planned Observing Strategy for the CDF-S observations.} 
\begin{tabular}{lrrr} \hline
{\em Filter}   & {\em
$T_{tot}$ (sec)} & {\em $N_{OB}$} & {\em $T_{exp}$ (sec) }  \\\hline  \hline
U    & 60000 & 13 & 900    \\
B    & 12000 & 8  & 300  \\
V    &  9000 & 6  & 300  \\
R    &  9000 & 6  & 300 \\
I    & 27000 & 9  & 600   \\ \hline \hline
\label{tab:strategy}
\end{tabular}
\end{center}
\end{table}
}

\begin{table}[h]
\begin{center}
\caption{Planned Observing Strategy for the CDF-S observations.} 
\begin{tabular}{lrrrrr} \hline
{\em Filter} & {\em $T_{exp}$ (sec)} & {\em $N_{exp}$} &  {\em
$T_{OB}$ (sec)}  & {\em $N_{OB}$}  & {\em $T_{tot}$ (sec)} \\\hline  \hline
U    & 900 & 5 & 4500 & 13 & 58500     \\
B    & 300 & 5 & 1500 & 8  & 12000  \\
V    & 300 & 5 & 1500 & 6  & 9000 \\
R    & 300 & 5 & 1500 & 6  & 9000 \\
I    & 600 & 5 & 3000 & 9  & 27000  \\ \hline \hline
\label{tab:strategy}
\end{tabular}
\end{center}
\end{table}

\begin{table}
\begin{center}
\caption{Nominal coordinates of OBs (see text for definition)}
\begin{tabular}{lcc} \hline\hline
{\em ID} & {\em $\alpha$(J2000)} & {\em $\delta$(J2000)}
  \\ \hline 
01  & 03:32:28  & -27:48:47  \\
02  & 03:32:34  & -27:48:17  \\
03  & 03:32:36  & -27:49:02  \\
04  & 03:32:32  & -27:49:32   \\
05  & 03:32:23  & -27:49:17   \\
06  & 03:32:21  & -27:48:31   \\
07  & 03:32:25  & -27:48:02  \\
08  & 03:32:30  & -27:47:47  \\
09  & 03:32:27  & -27:49:47  \\
10  & 03:32:19  & -27:50:02  \\
11  & 03:32:38  & -27:47:32  \\
12  & 03:32:40  & -27:50:17  \\
13  & 03:32:17  & -27:47:17  \\ \hline\hline
\label{tab:mainpos}
\end{tabular}
\end{center}
\end{table}

\begin{table}
\begin{center}
\caption{Relative offsets of each exposure in a OB}
\begin{tabular}{lcc} \hline\hline
{\em Dith. \#} & {\em $\Delta \alpha \cos \delta(\arcsec)$ }  & {\em
$\Delta \delta (\arcsec)$} \\ \hline
1 &   0  & 0  \\
2 & $-$70  & +21  \\
3 & $-$35  & $-$42  \\
4 & +35  & +42  \\
5 & +70  & $-$21  \\ \hline\hline
\label{tab:dither}

\end{tabular}
\end{center}
\end{table}

\begin{figure} 
\centerline{\hbox{\psfig{figure=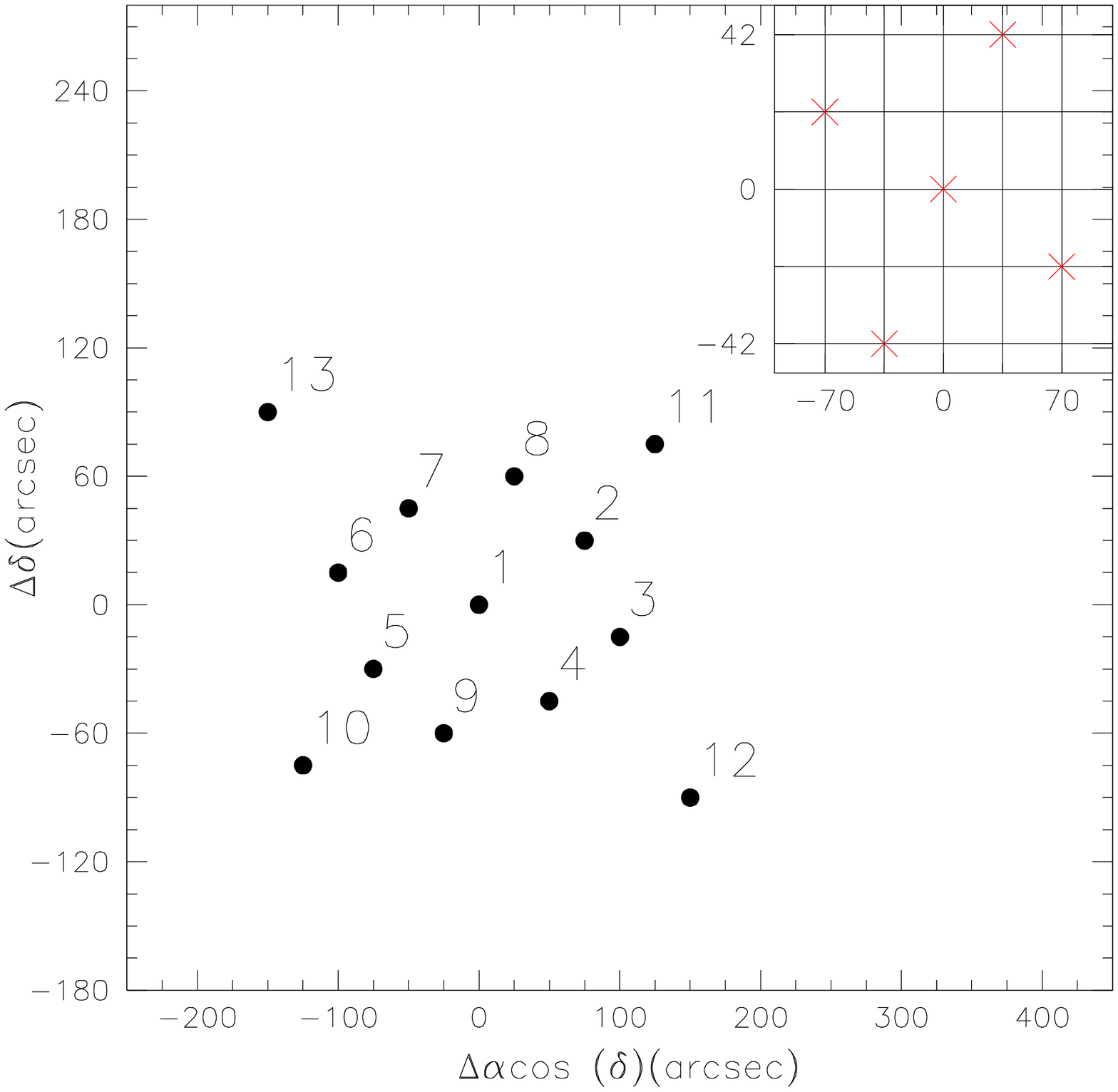,angle=0,width=8.8cm,clip=}}}
\caption{Schematic view of the dithering pattern used in the
      observations. Large filled circles mark the coordinates of each
      OB relative to the nominal center of the field (OB~\#1 in
      Table~\ref{tab:mainpos}). {\em (Inset)}: The crosses mark the
      offset, in arcsec, of each single exposure within an OB, with
      respect to its center.  }
\label{fig:pattern}
\end{figure}

The optical observations span six observing runs over a period of one
year (ESO observing periods 64 and 66), distributed over a total of 12
nights.  The log of the observations is summarized in
Table~\ref{tab:logs}. The table lists: in column (1) the date of the
observations; in column (2) the passband; in column (3) the
integration time; and in columns (4) and (5) the range and average
seeing during the night as measured by the seeing monitor.  Not listed
are the $B$-band observations taken during the commissioning phase of
the WFI, in January 1999, and the $U$-band data of the COMBO survey
(Wolf \etal 2000). Both data sets have been re-processed, for the sake
of homogeneity, and used to produce the final combined images.
Table~\ref{tab:obs} summarizes the data available for the CDF-S used
in this paper. The table lists: in column (1) the band; in column (2)
the total integration time; in column (3) the total number of frames;
and in column (4) the average seeing of the final co-added images.

\begin{table}
\begin{center}
\caption{Log of Observations.}
\begin{tabular}{lcrcc} \hline\hline
{\em Date}   &  {\em Filter} & {\em $T$ }  & {\em seeing range} &
{\em mean seeing}\\
             &               & (seconds)   &  (arcsec)  &  (arcsec) \\
 \hline
Nov 04 1999   & U     & 11700 & 0.62-0.97  &  0.77 \\
             & I     &  8400 & 0.77-1.71  &  1.04 \\ 
Nov 05 1999   & U     & 13500 & 0.74-0.94  &  0.82 \\ 
Nov 06 1999   & U     & 14400 &  0.48-0.83 &  0.64 \\ 
Nov 07 1999   & I     & 17700 & 0.76-1.26  &  0.97 \\ 
Nov 08 1999   & V     & 13500 & 1.40-1.81  &  1.61 \\ 
             & I     &  6000 & 1.20-2.08  &  1.64 \\
\hline
Dec 03 1999   & V     & 6000  & 0.46-0.98  &  0.66 \\ 
Dec 04 1999   & V     & 3000  & 0.48-0.72  &  0.58 \\ 
             & R     & 9000  & 0.47-0.79  &  0.58 \\ 
             & I     & 1800  & 0.67-0.77  &  0.71 \\
\hline
Oct 26 2000   & U$^{\prime}$   & 13000 & 0.57-1.04 &  0.78 \\ 
Oct 27-2000   & U$^{\prime}$   & 11700 & 0.52-1.01 &  0.70 \\ 
\hline
Nov 27 2000   & U$^{\prime}$   & 9000  & 0.50-0.76  & 0.59 \\ 
Nov 28 2000   & U$^{\prime}$   & 9000  & 0.67-1.06  & 0.87 \\ 
Nov 29 2000   & U$^{\prime}$   & 4500  & 0.66-1.88  & 0.76 \\
 \hline
\hline
\label{tab:logs}
\end{tabular}
\end{center}
\end{table}

\begin{table}
\caption{Summary of the CDF-S observations. }
\begin{threeparttable}
\begin{tabular}{cccc}
\hline\hline
Filter & $T_{eff}$ & $N_f$ & Seeing  \\ 
 & (sec) &  &  (arcsec)    \\
\hline
$U$    & 47600 & 55 & 0.98  \\ 
$U'$   & 43600 & 49 & 1.06  \\
$B$\tnote{a}  & 24500 & 30 & 1.02  \\ 
$V$    &  9250 & 33 & 0.97\\ 
$R$    &  9250 & 33 & 0.86 \\ 
$I$    & 26900 & 46 & 0.93 \\ 
\hline\hline
\label{tab:obs}
\end{tabular}
\begin{tablenotes}
\item[a] commissioning data
\end{tablenotes}
\end{threeparttable}
\end{table}

To assess the effectiveness of the adopted dithering
pattern Figure~\ref{fig:weights} illustrates the exposure time map
associated with the final $R$-band image which is an example of one of
the passbands with the shortest exposure and is thus the least
homogeneous (see Section~\ref{sec:results}). In this particular case
the flux ratio of the best sampled region to that at the center of the
image is a factor $\sim2$.  This is due in part to the clipping
($\sim$ 200 pixels) of the CCD frames not taken into account in the
original design of the dithering pattern. In the weight-map the
effects of at least three satellite tracks can be seen. It is
important to point out that some of the survey data have been obtained
using dithering patterns different from the one shown in
Figure~\ref{fig:weights} (\eg the $B$-images taken during the
commissioning of WFI). In fact, during the course of the survey the
pattern has been changed at least once to improve the uniformity of
the final image.

\begin{figure} 
\caption{Weight map corresponding to the final co-added $R-$band
image.  The bright regions correspond to locations in the image which
have higher sensitivity. While the effect of the inter-chip gaps are
minimized, their imprint can still be seen and correspond to the
darkest regions of the map (see text for more details). Also note the
slightly lower sensitivity in regions where satellite tracks have been
detected in individual frames and removed. At least two of them are
clearly seen on the upper part and on the right side of the weight map,
respectively.}
\label{fig:weights}
\end{figure}

The present optical survey complements both recent infrared
observations reported by Vandame \etal (2001) as well as
optical/infrared observations conducted as part of the EIS-DEEP survey
first released in 1998 and currently under revision (Rengelink \etal
1998).  The coverage of this field by these different sets of data is
shown schematically in Figure~\ref{fig:pointings}.

\section{Data Reduction}
\label{sec:reductions}

\subsection {Pipeline Processing}

The WFI images were processed using the refurbished EIS pipeline (\eg
da Costa \etal 1999; da Costa 2001) which includes new operational
modules to enable unsupervised pipeline reduction of WFI images as
well as new software routines for carrying out the basic steps of
image processing and calibration. Given the scope of the surveys being
carried out by EIS, special attention has been given to the
development of a survey software system to ensure the adequate and
timely processing of large amounts of data for long periods of time in
a uniform way. For instance, the complete DPS survey will consist of
some 2800 WFI science frames representing over 1 Terabyte of data, to
which should be added a comparable amount of data from the so-called
Pre-Flames survey not discussed here.  While a detailed account of the
extent of these changes is beyond the scope of the present paper, some
of the main features are briefly summarized below.

One of the main changes has been the complete removal of different
IRAF packages (\eg xccdred, drizzle) used earlier to process images
from CCD mosaics and in the co-addition phase. Instead, WFI images are
now split into individual chips at the very beginning of the
process. The main advantages of adopting this procedure are that the
pipeline uses a common library of routines, single and multi-chip
images are treated in the same way and that the EIS pipeline is more
easily adaptable to use PC-based Linux clusters for
parallel-processing images which will eventually lead to a higher
throughput.

\begin{figure} 
\caption{Schematic view of the coverage of the CDF-S by the EIS
program using different instruments. The outer white box delineates the
coverage of the present optical survey (WFI). The $2\times2$ black
mosaic and the two overlapping gray boxes delineate the area covered
by the infrared (SOFI) and optical (SUSI2) observations, respectively,
carried out by EIS-DEEP survey at the NTT. Finally, the $4\times4$
white mosaic represents the complete infrared coverage, including the
Deep Public Survey data presented by Vandame \etal (2001).  The
background image is taken from the DSS. }
\label{fig:pointings}
\end{figure}

A major effort has also been made to address specific problems
encountered with the data files coming from WFI. These have constantly
evolved which makes them difficult to handle from the point of view of a
pipeline. One of the main challenges has been the lack of information
in the FITS header regarding the reference pixel. This prevented the
unsupervised reduction of the WFI data due to frequent failures in
the astrometric calibration from run to run. As the EIS observations
with WFI span a long period of time (over two years), during which
several upgrades of the data acquisition were made, it was necessary
to develop robust procedures to cope with these various changes
automatically.

Using tools available from the multi-resolution visual model package
(MVM) developed by Bijoui and collaborators, a special routine was
developed to find a first-order astrometric solution for an image,
searching a large ($\sim6\times$) area surrounding the nominal
position of the pointing. The method uses one of the science frames in
the run and takes the available information in the image header
regarding the pointing only as a first guess. This image is quickly
reduced and decomposed using a wavelet transform. The lowest
resolution component is then used to cross-correlate features against
a low-resolution mock image representing the reference catalog over a
much larger area than the original image.  The low-resolution science
image is then moved around in a spiral pattern starting from its
nominal position as given in the image header.  At each step a
cross-correlation function between the features in the real and mock
images is computed and the process resumed. At the end of the search
the position with the largest amplitude of the cross-correlation is
used to define a first approximation for the location of the reference
pixel for that run.  This procedure is aimed at correcting only for
large shifts. A more refined astrometric calibration is performed at a
later stage.

Another important generic routine was developed to recognize different
type of exposures (calibration frames such as dome, sky flats and
bias, photometric standard stars or science frames), reject bad frames
and associate frames into pseudo-OBs to be reduced according to
pre-defined prescriptions depending on their nature and passband. The
association is based on the frame type and on the spatial separation
and time interval between consecutive frames.  Next, the removal of
instrumental signatures such as bias subtraction, flat-fielding and
de-fringing are carried out using identical procedures as those
applied for infrared images as presented by Vandame \etal (2001). Some
of the main features of the software worth mentioning are that the
stacking of images is carried out in pixel space (see below), thereby
allowing the optimal rejection of cosmic rays and bad pixels using
sigma--clipping.  Satellite tracks, quite frequent in wide-field
images, are efficiently masked on the individual exposures using the
Hough transform (Vandame 2001), thereby minimizing their impact on the
final stacked image. The number of satellite tracks removed in a
single passband ranges from 9 (\eg $R$-band images) to 75 ($U$-band
images) depending on the exposure time.  In principle, other linear
features such as diffraction spikes may also be treated in the same
way.  The net result of these new developments has been the production
of remarkably clean final images, significantly superior to those of
earlier releases. These results lead to the production of considerably
more reliable source detections since the successful masking of
features at the level of individual exposures dramatically minimizes
the number of false detections in the final image. Finally, using the
sky-flats a first evaluation of the relative gains of the different
chips is conducted for each passband which it is later applied to the
science frames.  These relative gains vary from passband to passband
and are, in some cases, especially in the UV, as large as 10\%.

The stacking of the images is done in two distinct steps. First, the
five images belonging to the same OB are warped to the first image of
the OB, which is used as reference for the pixel based relative
astrometry. All the images in the OB are then averaged. The second
step involves warping all the co-added images corresponding to each OB
to an absolute astrometric reference catalog (see next section). The
resulting images are then co-added taking into account their
respective weight-maps, noise and flux scales. For this particular
data set the warping of the images has been performed using the {\it
nearest-neighbor} approach. The advantage of this warping is that the
noise of the final image is uncorrelated. However, the astrometric
solution is less precise due to discreteness effects.  A new version
of the warping has been developed allowing for a suite of kernel
functions.

In contrast to other instruments (\eg FORS) the fringing observed in
$I$-band with WFI has a peak-to-peak amplitude of $\sim10\%$ relative
to the background and varies on scales of hours. The de-fringing is
performed in exactly the same way as for the infrared images, except
that the number of frames in an OB is considerably smaller, making a
good estimate harder. Currently the residual contribution after
de-fringing is comparable to the image noise. However, as in the case
of the infrared, it would be preferable to decrease the integration
time of each exposure and increase the number of dithered
images. Based on the current findings future $I-$band observations
will be done using a different observing strategy.

The implementations described above make the EIS pipeline a tool
capable of handling data from different telescope/instrument setups
with different implementations of the FITS header. Finally, it is
worth pointing out that using the available software several new
developments are planned for the future. These include seeing
deconvolution, morphological classification and difference image
photometry for variability studies.

\subsection{Astrometric calibration}

The astrometric calibration of the images was performed using as the
reference catalog a pre-release version of the {\em Guide Star
Catalog-II} (GSC-II). The GSC-II (McLean \etal~ 2001) is based on
multi-passbands all-sky photographic plate surveys including the
Palomar Observatory Sky Survey (POSS-I-II), the SERC and the ESO Red
Survey. The astrometric calibration of GSC-II has been obtained
relative to Hipparcos (Perryman \etal 1997) and Tycho catalogs
(H{\o}g~\etal 1997) and the {\em ACT} (Urban, Corbin \& Wycoff
1998). The astrometric calibration was done using the method developed
by Djamdji \etal (1993) based on a multi-resolution decomposition of
images using wavelet transforms (MVM).  An implementation of these
algorithms to stack images, referred to as MVM-astrometry, has been
done for EIS as is described above.  This package has proven to be
efficient and robust for pipeline reductions.

\subsection{Photometric Calibration}

Magnitudes were calibrated to the Johnson-Cousins system using Landolt
standards taken from Landolt (1992). While standard stars were
observed on every night, to optimize the observing time most
observations were used to monitor the zero-point of the night and to
compute color terms. For the final calibration, three nights in 
October 2000, all photometric, were used to obtain measurements of
standards over a broad range of airmasses. During these nights a set
of dithered images in each passband was taken overlapping two adjacent
pointings (Deep2-b and Deep-2c). These were then used to set the
absolute flux scale for all passbands, except the $I-$band due to
problems with de-fringing. For this case the night of Nov 04 1999 was
used to determine the $I$-band calibration.

Photometric solutions were obtained using the new EIS photometric
pipeline which is fully integrated to the associated database. The tool
provides a convenient environment to obtain, verify and store the
results of the fits. In the interactive mode it is possible to fit the
data after rejecting individual measurements, stars and
chips. Alternatively, one can also easily obtain independent solutions
for each chip. Unfortunately, the number of standards for the fields
considered does not allow independent solutions for each of the eight
chips. Therefore, the present calibration had to rely on solutions
including all the available chips.

Magnitudes for Landolt stars were obtained using an aperture 6~arcsec
in diameter, which proved to be adequate, by monitoring the growth
curve of all the measured stars.  The accuracy of the zero-points are
estimated to be in the range $\pm$0.03~mag to $\pm$0.08~mag, with the
$U$-bands having the largest uncertainties. At face value these
results indicate that the gain correction applied to the different
chips based on the flatfield exposures is adequate for most passbands
with the possible exception of the $U$-band. A final determination of
the relative gains will require a more careful monitoring of the
amplitude of these corrections and chip-based solutions.  A set of
secondary stars for all the Landolt fields considered would also be
extremely useful to always allow a chip-based calibration. Efforts in
this direction are already underway by several groups.

The computed extinction coefficients compare well with those estimated
from the mean extinction curve of the Chilean sites. Finally, the
following color terms, relative to the Johnson-Cousins (JC) system,
have been empirically determined:

\medskip
$
\begin{array}{l}
(U_{JC}-U'_{EIS})= 0.04     \times(U-B)_{JC},\\
(U_{JC}-U_{EIS})= 0.03      \times(U-B)_{JC},\\
(B_{JC}-B_{EIS})= 0.14      \times(U-B)_{JC},\\
(B_{JC}-B_{EIS})= 0.24      \times(B-V)_{JC},\\
(V_{JC}-V_{EIS})= $--$0.11  \times(B-V)_{JC},\\
(V_{JC}-V_{EIS})= $--$0.21  \times(V-R)_{JC},\\
(R_{JC}-R_{EIS})= $--$0.04  \times(V-R)_{JC},\\
(R_{JC}-R_{EIS})= $--$0.04  \times(R-I)_{JC},\\
(I_{JC}-I_{EIS})= 0.25      \times(R-I)_{JC},\\
\end{array}
$
\medskip

These equations can be inverted to a more useful form, expressing the
color terms as a function of the observed colors yielding:

\medskip
$
\begin{array}{l}
(U_{JC}-U'_{EIS})= 0.04 \times(U'-B)_{EIS},\\
(U_{JC}-U_{EIS})= 0.03 \times(U-B)_{EIS},\\
(B_{JC}-B_{EIS})= 0.36  \times(B-V)_{EIS},\\
(V_{JC}-V_{EIS})= $--$0.08  \times(B-V)_{EIS},\\
(R_{JC}-R_{EIS})= $--$0.05  \times(V-R)_{EIS},\\
(I_{JC}-I_{EIS})= 0.35 \times(R-I)_{EIS},\\
\end{array}
$
\medskip

These values have a typical error of $\pm 0.03$ and are in good
agreement with those computed using the response function of the
filters considered, except for $U'$ for which the difference in the
color term can be $\sim0.2$ mag. This is because the UV-response of
the system is at the time of writing still poorly determined.

The magnitudes were also corrected for galactic absorption, using
$E(B-V)=0.0089$ as derived from Schlegel, Finkbeiner \& Davis (1998),
yielding $A_U=0.05$~mag, $A_B$=0.04~mag, $A_V$=0.03~mag,
$A_R$=0.02~mag and $A_I$=0.02~mag. All magnitudes have been converted
to the $AB$ system, unless otherwise specified, using the following
relations: $U'_{AB} = U' + 1.04$; $U_{AB} = U + 0.80; B_{AB}= B -
0.11; V_{AB}= V; R_{AB}= R + 0.19;$ and $I_{AB}=I + 0.50$.

\subsection{Image Products}

The images being released are fully astrometrically (conical
equal-area projection; COE) and photometrically calibrated (normalized
to 1~sec exposure time).  The FITS files being provided include both
the pixel and weight maps as FITS extensions.  A stand-alone software
tool to convert the COE images to best-fit TAN projection by modifying
the image headers is available through the EIS web pages. The
astrometric information is stored in the world coordinate system (WCS)
keywords in the FITS headers. Similarly the photometric calibration is
provided by the zero-point and its error available in the header
keywords ZP and ZP\_ERR. The zero-point includes the normalized
zero-point of the photometric solution and the atmospheric extinction
correction.  In the header one can also find a product identification
number (P\_ID}) which should always be used as reference. The headers
also provide information on the number of stacked frames and the
on-source total integration time.  Additionally, the seeing obtained
by measuring the FWHM of bright stars in the final stacked images is
stored in the header keyword SEE\_IMA.

\void{
\begin{figure*}
\hspace{0.2cm}
\caption{An example of the final coadded image (left panel) and the
corresponding weight map (right panel). The lines seen in the weight
map are caused by the astrometric re-mapping of the frames.}
\label{fig:imageK}
\end{figure*}
}

High-resolution images in the different passbands can be found at the
URL ``http://www.eso.org/science/eis/''.  The image quality of the
final images can be assessed by investigating the nature and the
amplitude of the PSF distortions. These are measured by using the two
components of the so-called polarization vector. These components are
derived from the major- (A) and minor-axis (B) and the position angle
($\theta$) as measured by SExtractor for point-like sources using the
following expressions:

$$e_{1} = \frac{A-B}{A+B}\;\mathrm{cos}(2 \theta) $$
$$e_{2} = \frac{A-B}{A+B}\;\mathrm{sin}(2 \theta)$$

\begin{figure} 
\centerline{\hbox{\psfig{figure=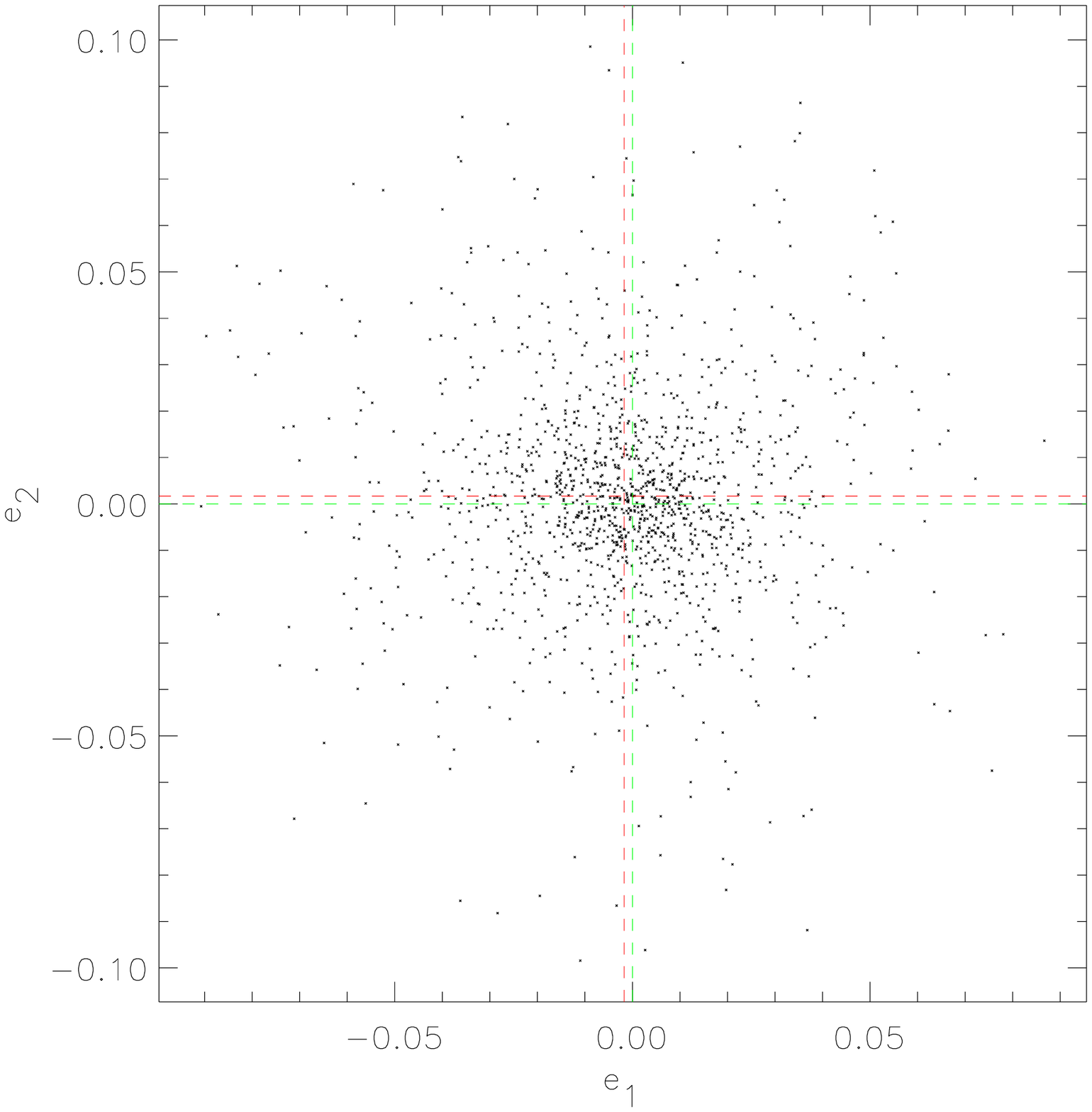,angle=0,width=8.8cm,clip=}}}
\caption{Amplitude of the PSF distortion over the entire final
$R$-band image. One line is centered on (0,0), the other on the actual
barycenter of the datapoints on ($-$0.002, 0.002) }
\label{fig:anisotropy}
\end{figure}

\begin{figure} 
\centerline{\hbox{\psfig{figure=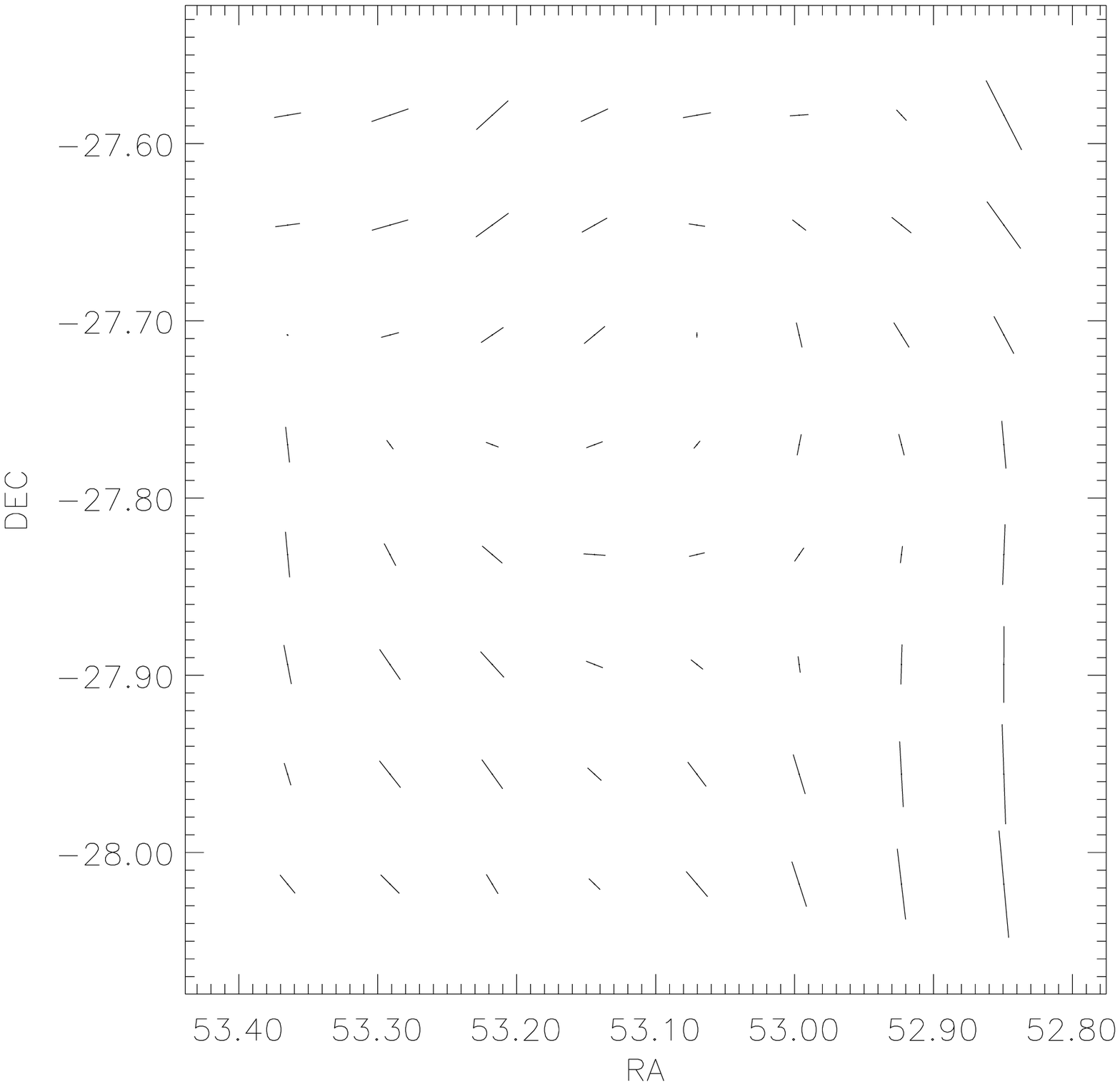,angle=0,width=8.8cm,clip=}}}
\caption{Vector representation of the PSF distortions for the final
$R-$band image smoothed over a scale of 7~arcmin. RA and DEC are in
degrees. The amplitude of the vectors are scaled relative to the thick
line segment shown on the upper left corner, which corresponds to an
amplitude of 0.0015.}
\label{fig:distortion}
\end{figure}

The quantities $e_{1}$ and $e_{2}$ give the amplitude of the
distortion. As an illustration, Figures~\ref{fig:anisotropy} and
\ref{fig:distortion} show the results obtained from the final $R$-band
image, which consists 30 co-added WFI
exposures. Figure~\ref{fig:anisotropy} shows the distribution of the
amplitude of the components, $e_{1}$ and $e_{2}$, for the individual
point sources observed over the entire field and in
Figure~\ref{fig:distortion} the polarization vector field smoothed
over a scale of 7~arcmin. The scale of the amplitude of the vectors is
set by the length of the small thick line shown on the upper left
corner of the figure, which corresponds to an amplitude of 0.0015
(0.15\%). These results attest to the excellent quality of the images
produced by WFI leading to a uniform and small PSF distortion
throughout the frame with an $rms$ amplitude $\lsim 2\%$.  Similar
results are obtained for the other passbands.

\section{Source Lists}
\label{sec:catalogs}

\begin{table*}
\caption{First 40 entries of the CDF-S $R$-band source list (all
magnitudes are given in the $AB$ system).}
\centerline{\hbox{\psfig{figure=
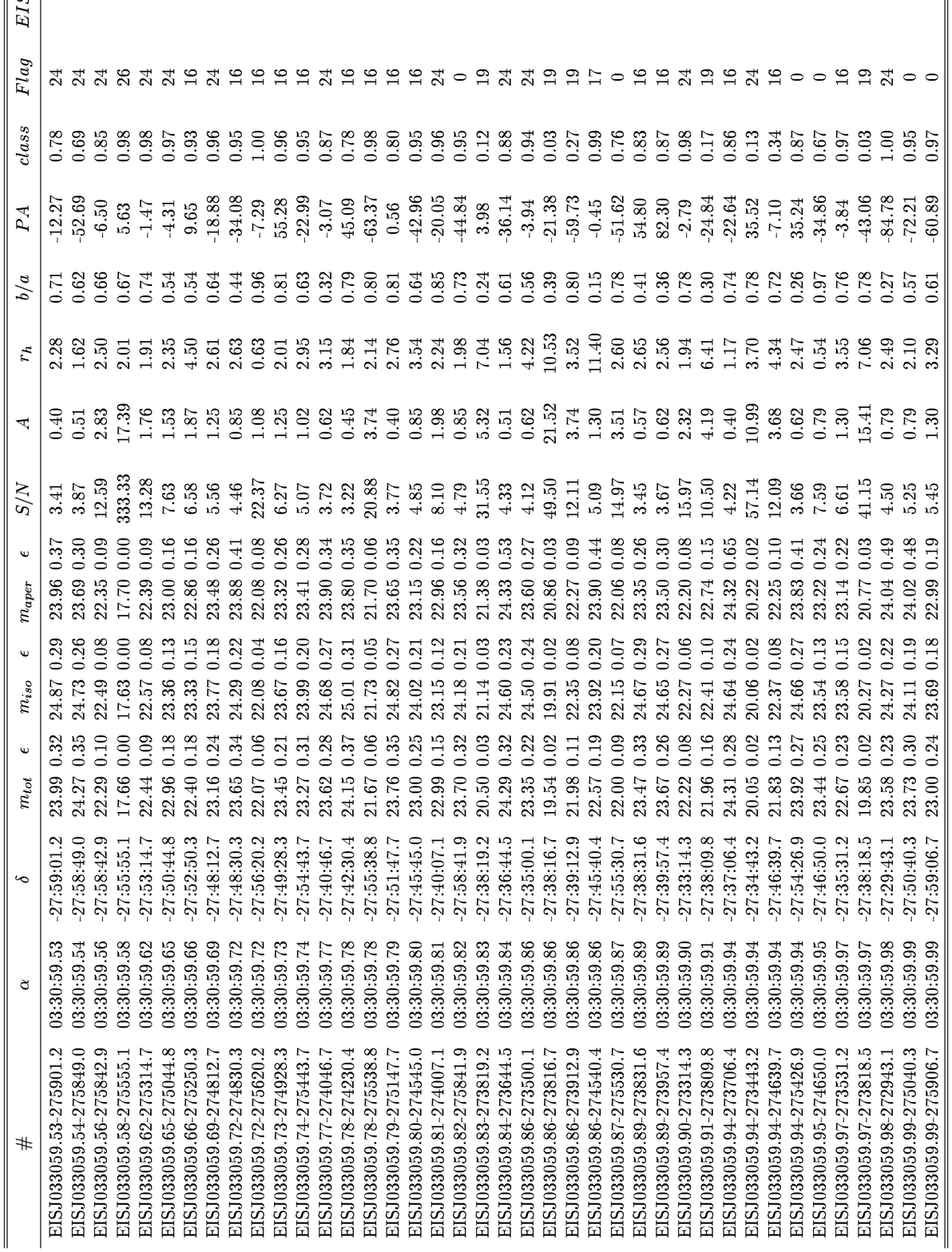,angle=0,width=\textwidth,clip=}}}
\label{fig:catalog}
\end{table*}

Source extraction was performed using the latest version of SExtractor
software (Bertin \& Arnouts 1996; ver. 2.2.1). This version resolves a
subtle problem, first identified in the prompt release of the Pilot
Survey, which caused the creation of regions devoid of sources when
applying SExtractor to WFI frames with weights
(Bertin~2001). Detection was carried out separately using the co-added
image of each passband and field.  The main parameters in the
detection are the smoothing kernel, taken to be a Gaussian with a FWHM
equal to 0.8 of that of the PSF measured on the frame; the minimum
number of connected pixels above the detection threshold, depending on
the seeing taken to be between 8 and 12 pixels, and the SExtractor
detection threshold, taken to be between 0.55 (for $U$) and 0.7 ($R$
and $I$) depending on the seeing. As an illustration, the tabulation
of the first 40 entries in the $R$-band source catalog is presented in
Table~\ref{fig:catalog}. All magnitudes are given in the $AB$ system.
The table lists:

Column (1): the EIS identification name

Columns (2) and (3): right ascension and declination (J2000.0);

Columns (4)-(9): total, isophotal and aperture (3~arcsec diameter)
magnitudes and respective errors. The first two magnitudes correspond
to the {\tt mag\_auto} and {\tt mag\_iso} magnitudes measured by
SExtractor.  The magnitudes have been corrected for Galactic
extinction taken from Schlegel \etal (1998). The errors are those
estimated by SExtractor and include only the shot-noise of the
measured source and background counts. Only objects detected with
signal to noise $S/N\geq3$ (based on the isophotal magnitude errors)
are included.

Column (10): an estimate of the $S/N$ of the detection, from the inverse
of the errors estimated for the isophotal magnitude;

Columns (11): the isophotal area $A$ of the object in square arcsec;

Column (12): the half-light radius $r_h$  in arcsec;

Column (13) and (14): minor to major-axis  ratio and the position angle;

Column (15): the stellarity index computed by SExtractor;

Column (16): SExtractor flags (see Bertin 1998)

Column (17): EIS flags; these flags are used to identify objects
in regions with very low weight or close to bright objects. Objects
detected in regions with weight $\geq30\%$ of the maximum weight
(proportional to the total integration time) and not affected by
bright stars have EIS flag=0. Objects with weight $<30\%$, located
at the edges of the frame, have EIS flag=1. Finally EIS flag=2
indicates that the object is located in a region masked out due to the
presence of a bright object.

Similar tables for the other passbands are available upon request in
ASCII format. Catalogs for the infrared mosaic and optical/infrared
color catalogs will be distributed as soon as they become available.
Note that besides distributing catalogs in tabular form as done here,
catalogs in FITS formats will become part of the distribution of final
products. From the inspection of the $R$-band weight map (see
Figure~\ref{fig:weights}) one sees that in contrast to catalogs
extracted from single chip detectors, describing source catalogs drawn
from multi-passband dithered observations of CCD mosaics requires
significantly more information than can reasonably be provided in a
single ASCII catalog. While the latter is adequate if the derived
objects list is to be used for optical cross-identification, it is
clearly not sufficient if one is interested in drawing statistical
samples with well-defined limiting magnitudes and color coverage. To
cope with these deficiencies future releases will be made using FITS
tables which will include: a FITS header with all the relevant
information about the observations (pointing, filter, instrument), a
FIELDS table which will include: the SExtractor parameters; the
magnitude and stellarity index chosen for the star/galaxy
classification; magnitude corrections (\eg AB, extinction) appropriate
for the filters used; a variety of limiting magnitudes (point source,
turnover, completeness, spurious objects) including the brightest
limiting magnitude to extract a homogeneous magnitude-limited sample;
the effective area corresponding to this sample taking into account
the area removed at the outer edges of the image based on the
weight-map and the masks around bright objects both now being produced
automatically by the pipeline according to user-specified
parameters. At the same time no objects will be removed but will carry
both their SExtractor and EIS flags, with the vertices of the
corresponding masks also being included in the FITS catalogs in a
specific fields table. This will also apply to color catalogs which,
in addition to information similar to single passband catalogs, must
also include others such as the definition of color context to clearly
indicate the area of intersection of images in different passbands. In
summary, the goal is to produce standard catalogs from which more
complex analysis can be carried out with or without the images and
associated weight maps.

The output of the new standardized procedures for single passband
catalog production is shown in Figure~\ref{fig:masks} for the catalog
extracted from the final $R-$band image. The figure shows the outer
bounding box defined based on the percentage of the peak exposure time
one wants to consider and the masks around saturated or bright
objects. Both the magnitude of the objects to be masked and the size
of masks can be set by the user. In this example one sees that not all
possibly spurious sources can be eliminated automatically. In
particular, in the lower right side of the figure ($RA=53.35$,
$Dec=-27.9$ degrees) one sees the effect of a ghost image of a bright
star which produces a ring-shaped excess of objects. While attempts
are currently being made to identify ghosts and automatically
eliminate these regions it is clear from the diversity of features
that in practice, further by-hand clipping of the image is required
depending on the specific application of the catalog. Tools for doing
so are also implemented in the pipeline with the results being stored
in the FITS catalog.

Each catalog produced by the pipeline goes through an automatic
verification process which carries out several tests to check the
reliability of the products by comparing them with other empirical
data and model predictions.  The results of some of these tests are
described in the next section.

\begin{figure*}
\caption{Projected distribution of galaxies extracted from the $R$-band
image. The figure shows the adopted frame and the masks automatically
produced by the pipeline around bright objects which are used to set the
EIS flags. It is important to emphasize that these parameters can be
re-defined.} \label{fig:masks} \end{figure*}

\section{Discussion}
\label{sec:results}

\subsection{Astrometry and Photometry}

\begin{figure}
\centerline{\hbox{\psfig{figure=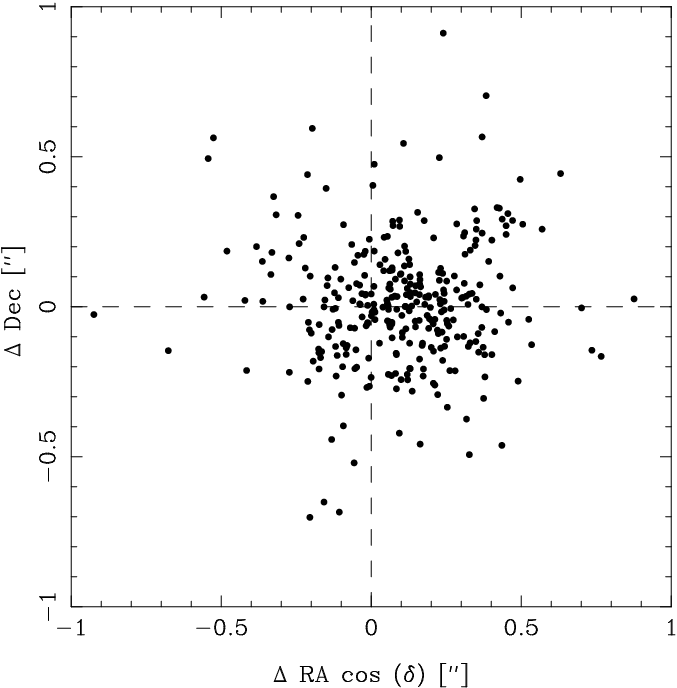,angle=0,width=\columnwidth,clip=}}}
\caption{Comparison in the $R$-band between the coordinates of objects in common with the
2MASS survey, offsets are computed as $EIS-2MASS$.}
\label{fig:astro_2mass}
\end{figure}

In order  to externally    verify  the accuracy of    the  astrometric
calibration the source  lists  extracted from the different  passbands
were compared with  that available from  the 2MASS survey (Cutri \etal
2000).  The number of objects  found in common  ranges from 286 to 415
depending   on   the passband   considered.  Using   ``good'' objects,
considering the various flags, in both catalogs the relative positions
of coincident pairs   were  computed  and   the  distribution of   the
differences  in    the  case    of   the  $R$-band    is   shown    in
Figure~\ref{fig:astro_2mass}. One finds an offset of about 0.12~arcsec
in right ascension, a negligible shift  in declination and an $rms$ in
the differences   of $\lsim0.23$~arcsec, independent  of the  passband
considered.   This  shows that    the  astrometric solutions   for the
different passbands are essentially identical  and an overall accuracy
of $\sim 0.16$~arcsec,  probably limited to  the internal  accuracy of
the reference catalog, is achieved. Note that the shift of the optical
data relative to 2MASS is smaller than that obtained with the infrared
data.  This, however, does not imply any inconsistency in the relative
astrometry of the infrared and optical EIS data but rather it reflects
the  problem described in Vandame  \etal (2001) with the astrometry of
the  pre-release of  GSC-II.  This  problem  will  be  fixed with  the
official release of the GSC-II. 

\begin{figure}
\centerline{\hbox{\psfig{figure=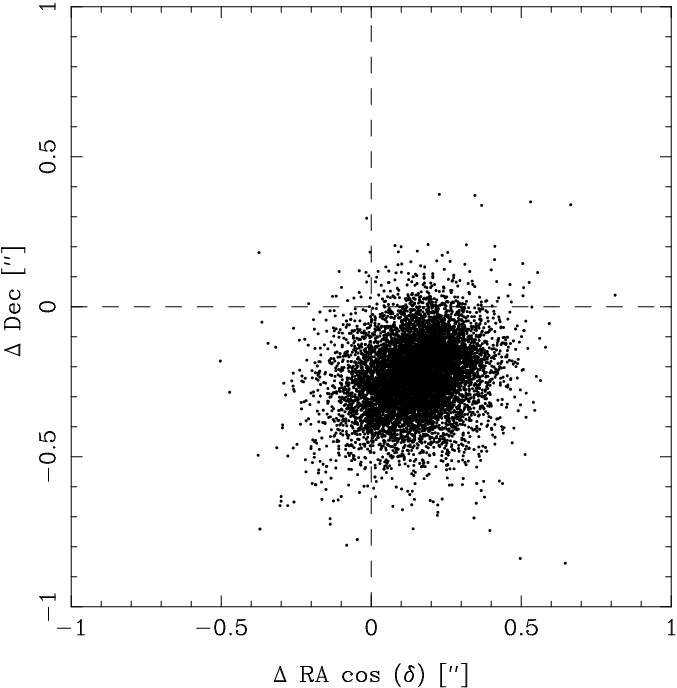,angle=0,width=\columnwidth,clip=}}}
\caption{Comparison between the location of objects in common with the
COMBO survey, offsets are computed as $EIS-COMBO$.}
\label{fig:astro_wolf}
 \end{figure}

An independent check of the astrometry can also be carried out by
comparing the present astrometry with that determined by Wolf \etal
(2000; COMBO survey) who recently obtained deep exposures of the same
field in different passbands using the same instrument. In particular,
their $R$-band data reaches $R=26.0$ (Vega) at $5\sigma$ for a total
integration time of 23700~sec. Within the overlapping area of both
surveys, there are: $\sim36800$ objects in common; $\sim$ 17500
detected only by COMBO; $\sim$ 4400 only by EIS; and $\sim$ 4200 have
multiple associations, which are not considered below.
Figure~\ref{fig:astro_wolf} shows the result of the comparison of the
positions for about 8500 objects in common. These objects include only
objects with stellarity index $>$0.9 in the COMBO $R$-band source list
and satisfying $R_{EIS}<25$~mag and $S/N (EIS)>$4. From this
comparison one finds a relative offset of 0.15~arcsec in right
ascension and $-$0.24~arcsec in declination, but more importantly a
remarkably small $rms$ of 0.12~arcsec in both directions, possibly
suggesting an accuracy of $\lsim0.10$~arcsec for each individual
catalog. This is well within the requirements for slit/fiber
positioning, a top requirement for the public survey. These results
are insensitive to the exact way the sample is chosen.

The COMBO data also enables a comparison of the photometric
calibration. Figure~\ref{fig:phot_wolf} shows the results of this
comparison in the Vega system from which one finds a zero-point offset
of $\sim-0.01$ with an $rms$ of about 0.01~mag for $16\lsim R_{EIS}
\lsim 20$~mag, while for objects brighter than $R\sim16$~mag
saturation sets in. For magnitudes fainter than $R\sim 20$ the scatter
increases and beyond $R\sim22$ one sees the Malmquist bias effect.

\begin{figure}
\caption{Comparison between the measured $R$ magnitudes of objects in
common with the COMBO survey as a function of the $R_{EIS}$ magnitude
(Vega system).}
\label{fig:phot_wolf} \end{figure}

\begin{figure}
\resizebox{0.52\columnwidth}{!}{\includegraphics{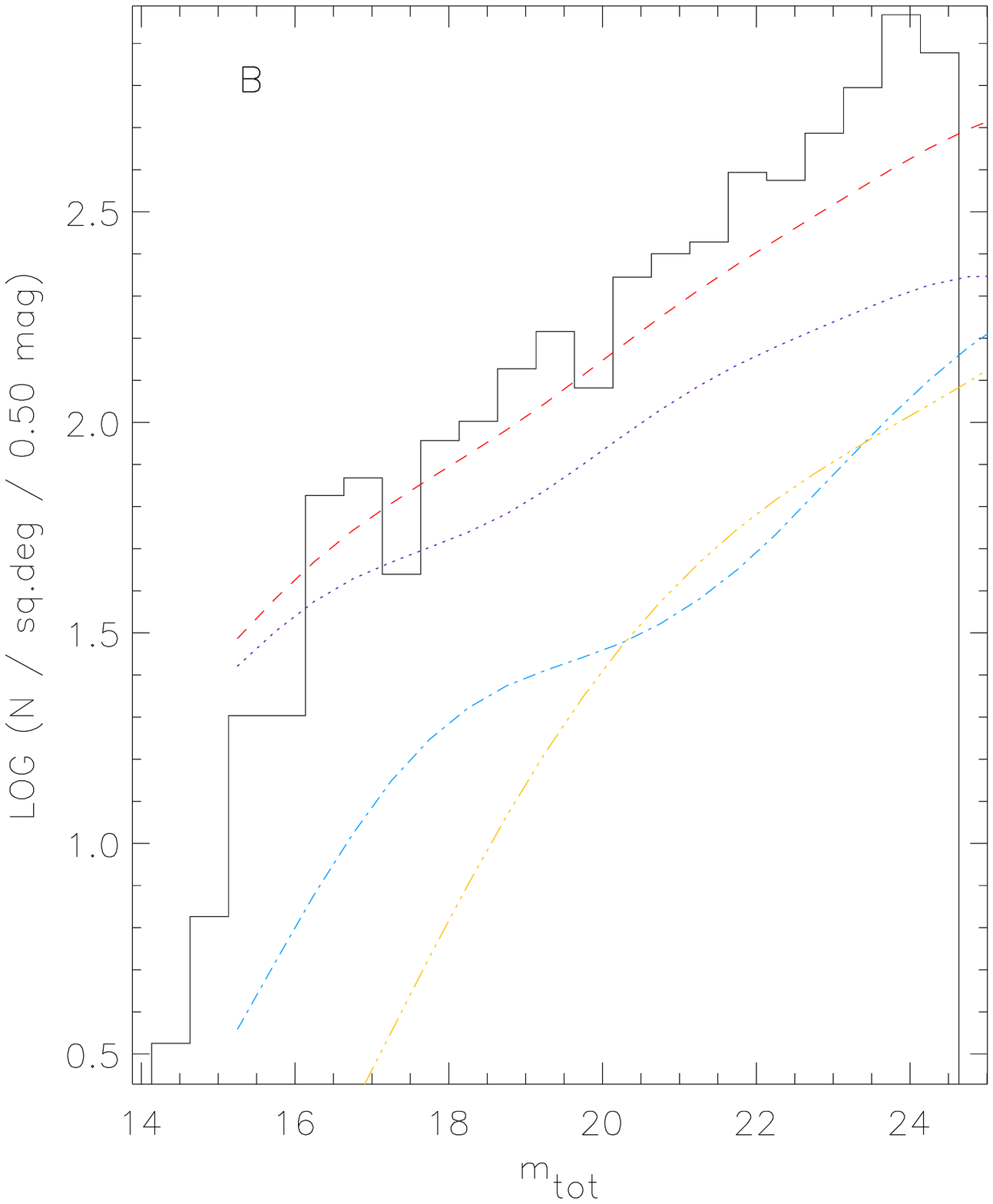}}
\resizebox{0.52\columnwidth}{!}{\includegraphics{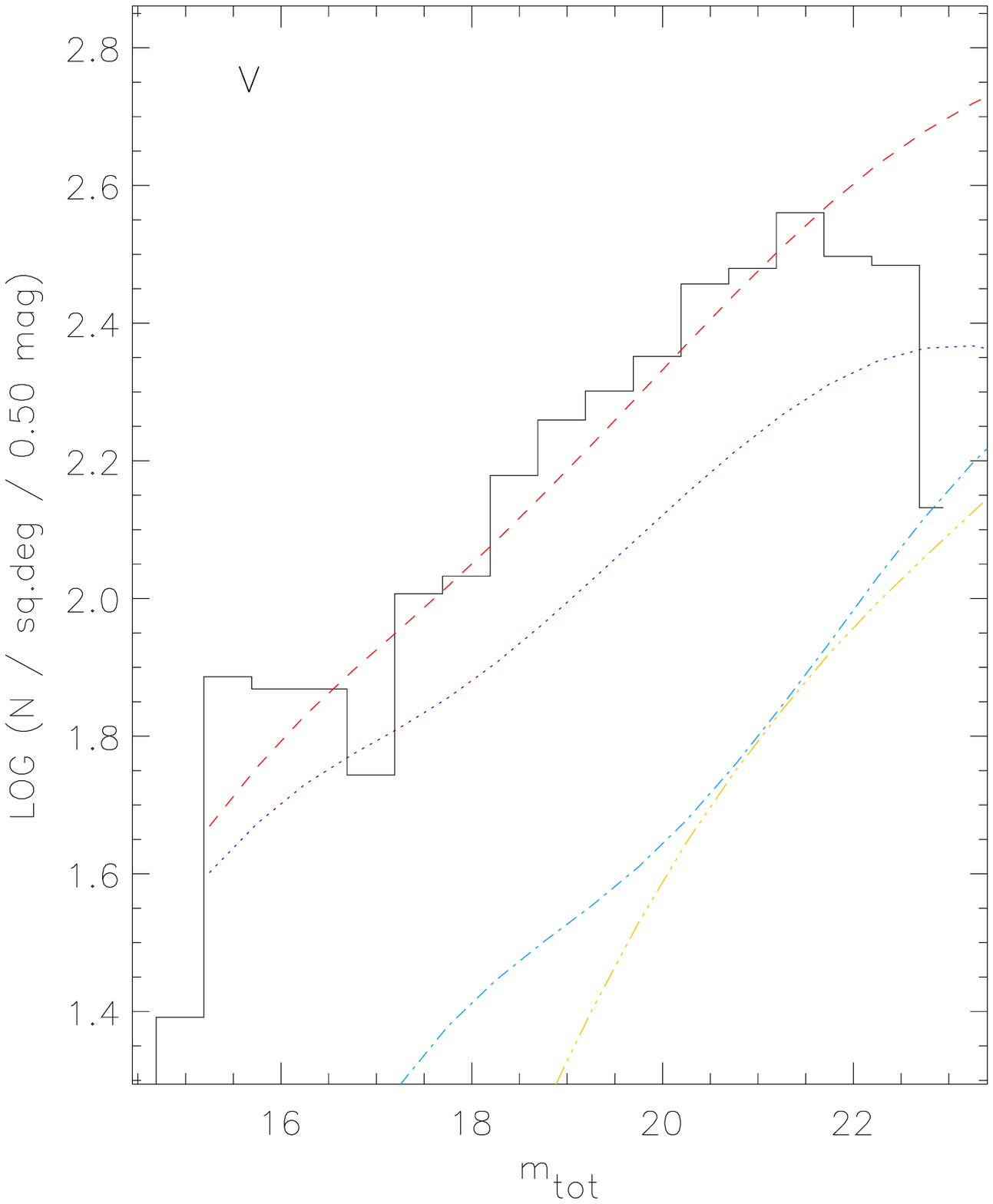}}
\resizebox{0.52\columnwidth}{!}{\includegraphics{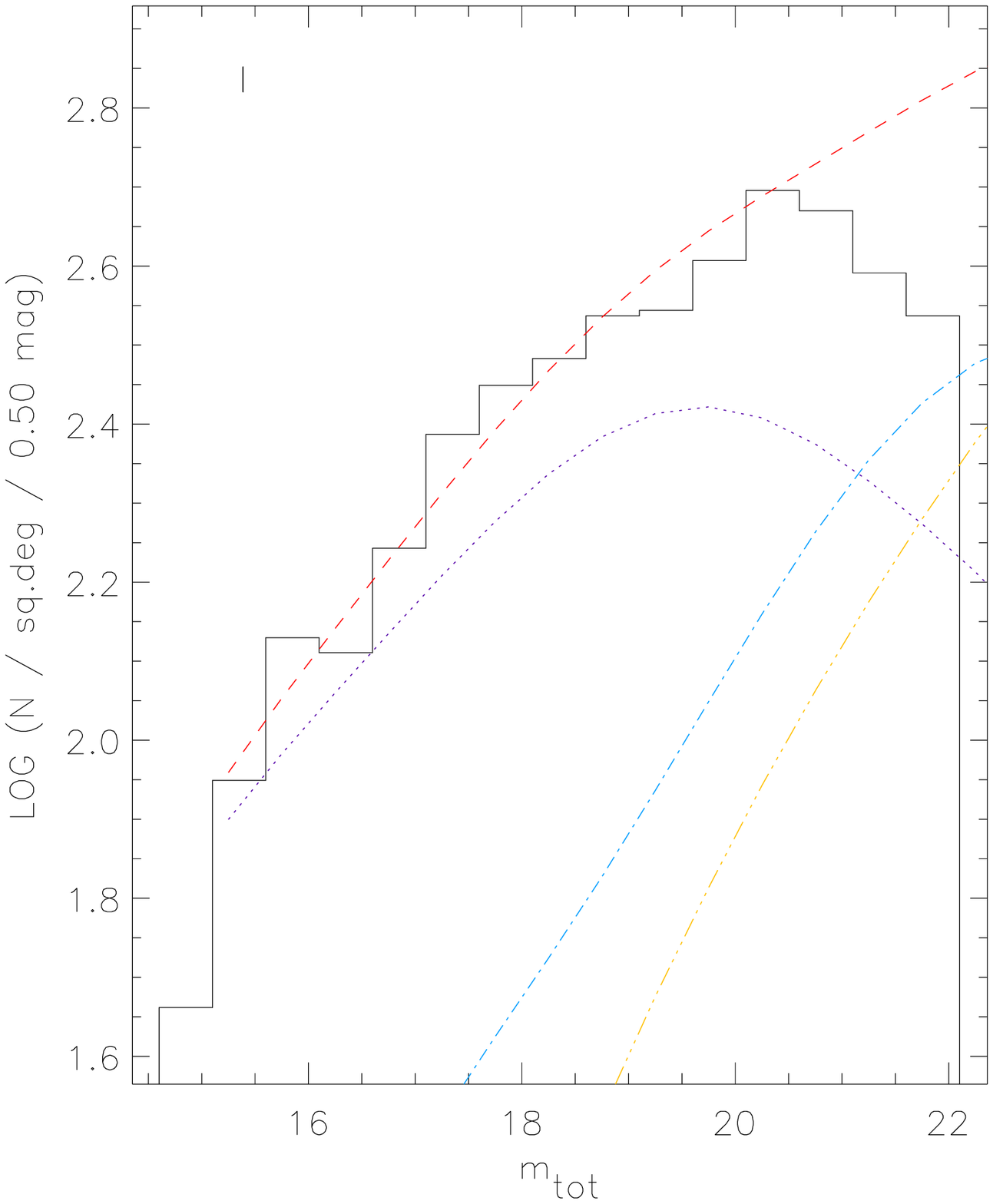}}
\caption{Comparison of stellar number counts in $B$ (top panel), $V$
(middle panel) and $I$ (lower panel) passbands obtained from the
present work (histogram) and those based on the Galactic model of
Mendez \& van Altena (1996; long-dashed line). In this plot, the
Magnitudes are in the Vega system.  The model includes an old disk
population (short dashed line), a thick-disk component (dot-dashed
line) and a halo component (dot-dot-dashed line).}
\label{fig:stcountsb}
\end{figure}

\subsection{Number Counts}

A simple statistic that can be used to verify the overall 
characteristics of the derived catalogs and the parameters chosen for
the star/galaxy classification is to compare the number counts of
stars and galaxies with model predictions and other empirical data.

Figure~\ref{fig:stcountsb} show a comparison of the stellar counts as
a function of the total magnitude (Vega system) with those obtained
using the Galactic model described by Mendez \& van Altena (1996),
using the standard parameters described in their Table~1. Considering
that no attempt has been made to choose model parameters to fit the
observed counts the agreement between observations and the model
predictions is very good. In the $B$-band, where the present survey is
deepest, there seems to be an excess of stellar objects. This is
possibly due to QSOs, which are estimated to have a surface density of
120 objects per square~degree per 0.5 mag interval at $B$=24
(Hatziminaoglou \etal, 2001).  This effect is not expected
in $V$ and $I$ because the surface density of QSOs at the magnitude
where the stellar counts drop is expected to be small, of order 30-40
per square~degree per 0.5 mag interval. Currently, the comparison of
the star counts is only possible for the three passbands
considered. However, this problem is being addressed and soon a more
generic galactic model will be available in the EIS pipeline to
compare the predicted counts in all passbands using the transmission
curves for the filters being used in a given survey setup (Girardi
2001).

\begin{figure*}
\resizebox{0.5\columnwidth}{!}{\includegraphics{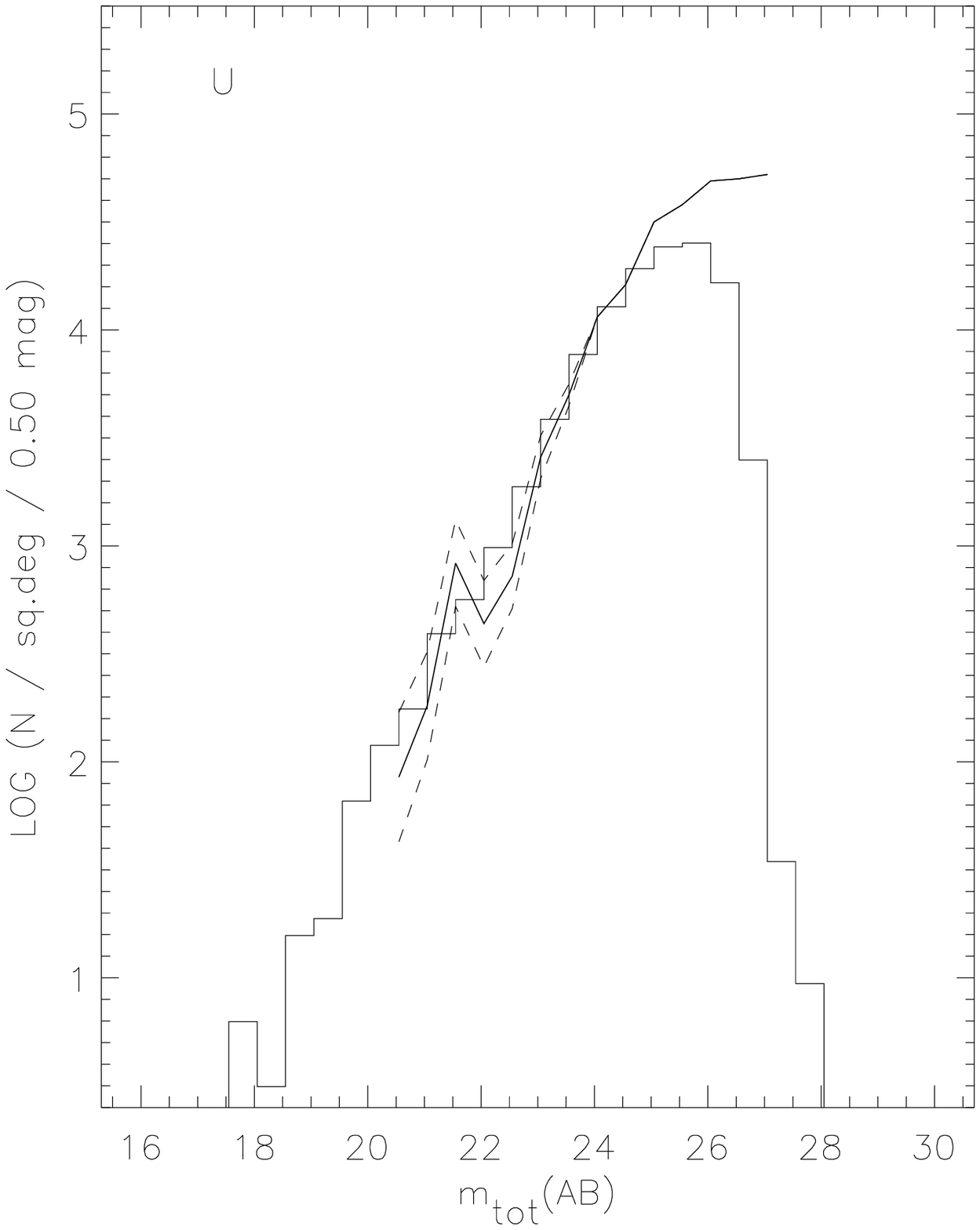}}
\resizebox{0.5\columnwidth}{!}{\includegraphics{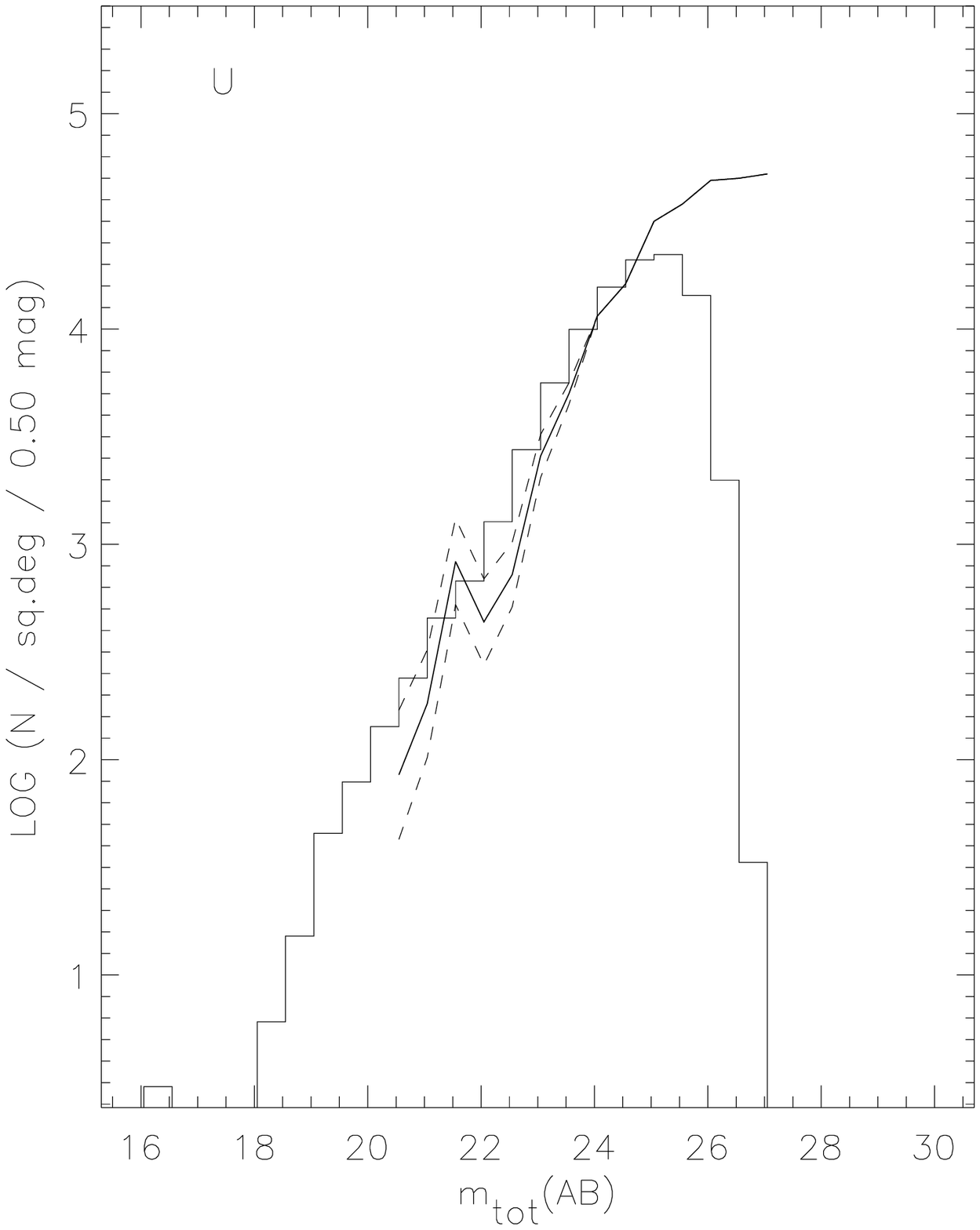}}
\resizebox{0.5\columnwidth}{!}{\includegraphics{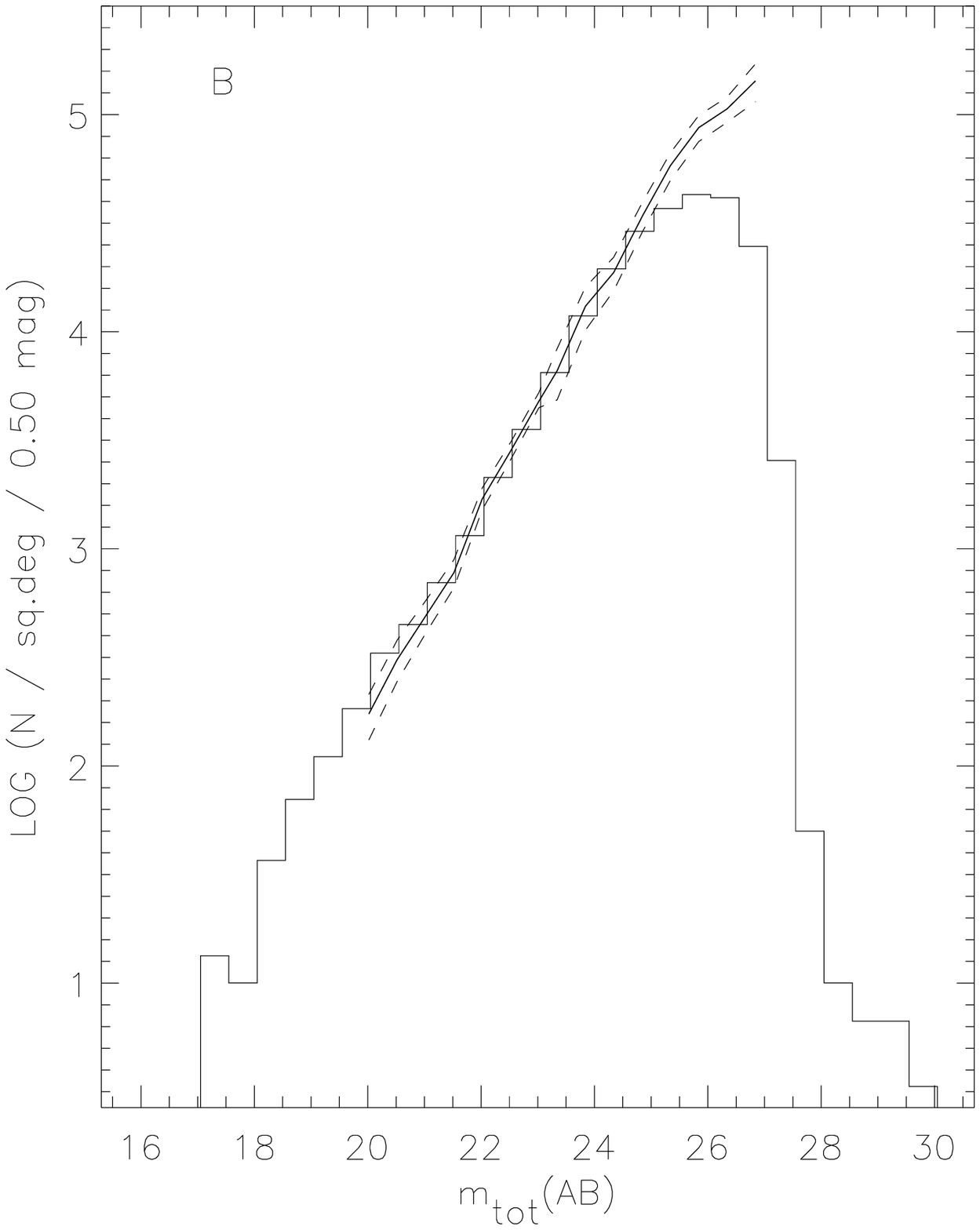}}
\resizebox{0.5\columnwidth}{!}{\includegraphics{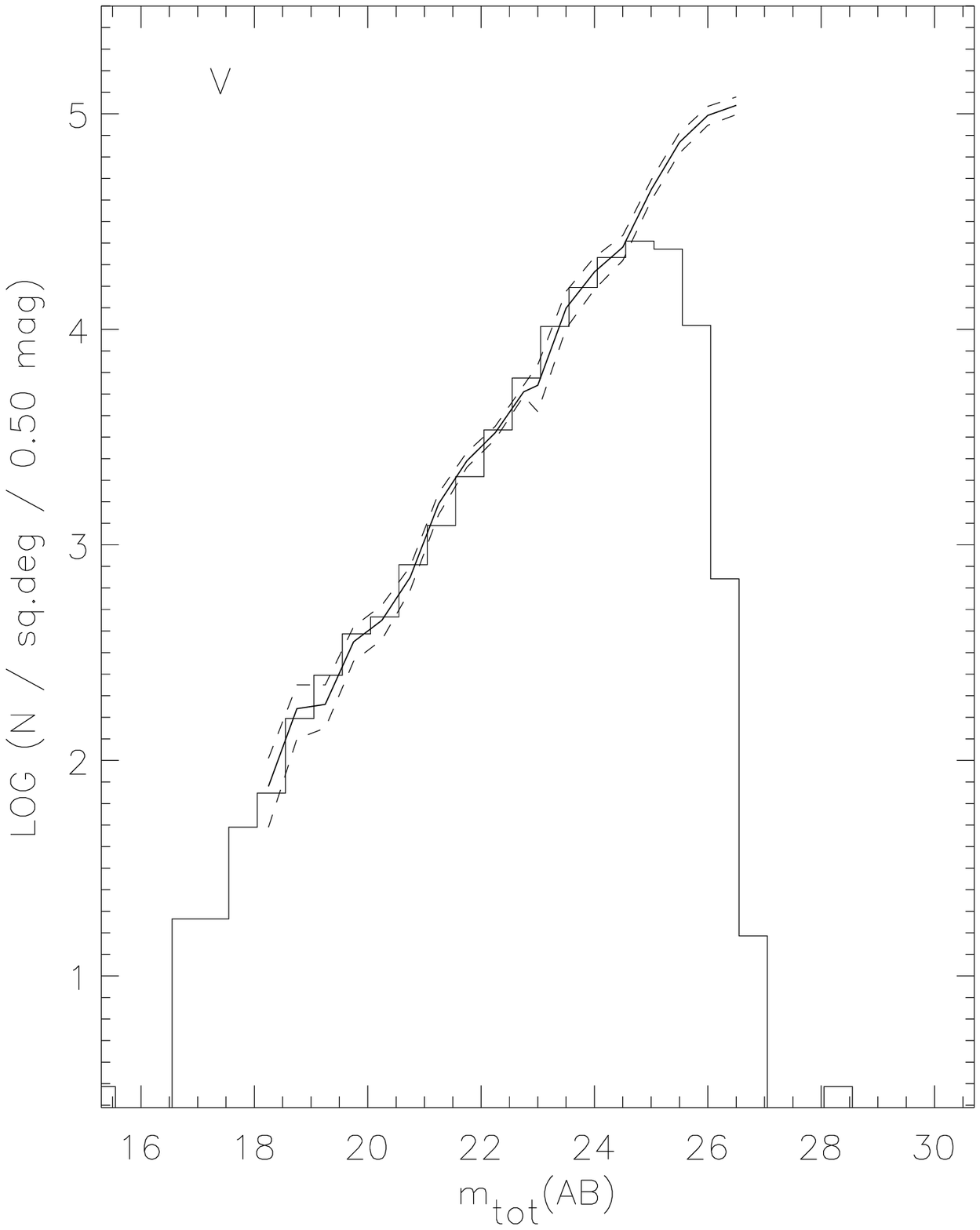}}
\resizebox{0.5\columnwidth}{!}{\includegraphics{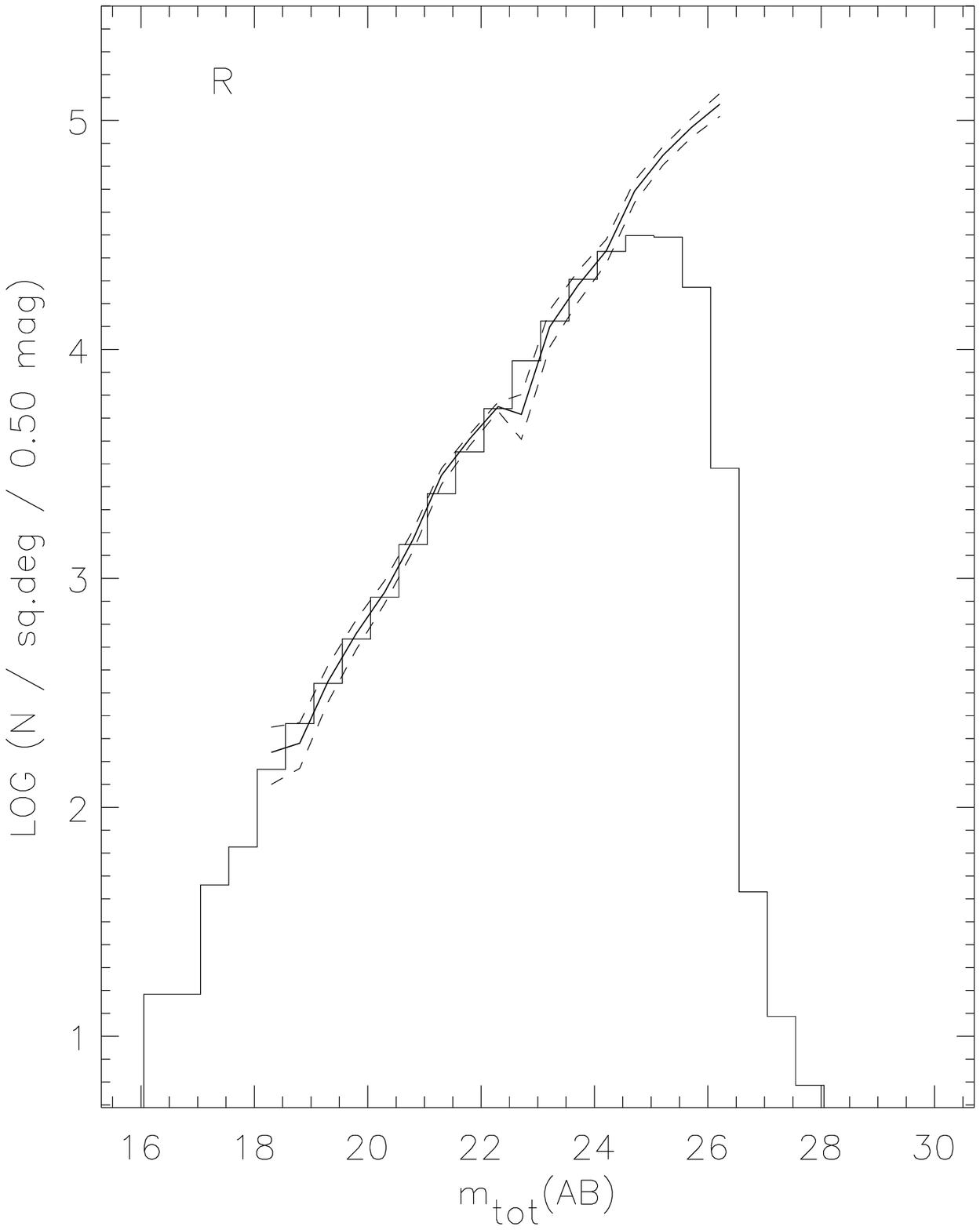}}
\resizebox{0.5\columnwidth}{!}{\includegraphics{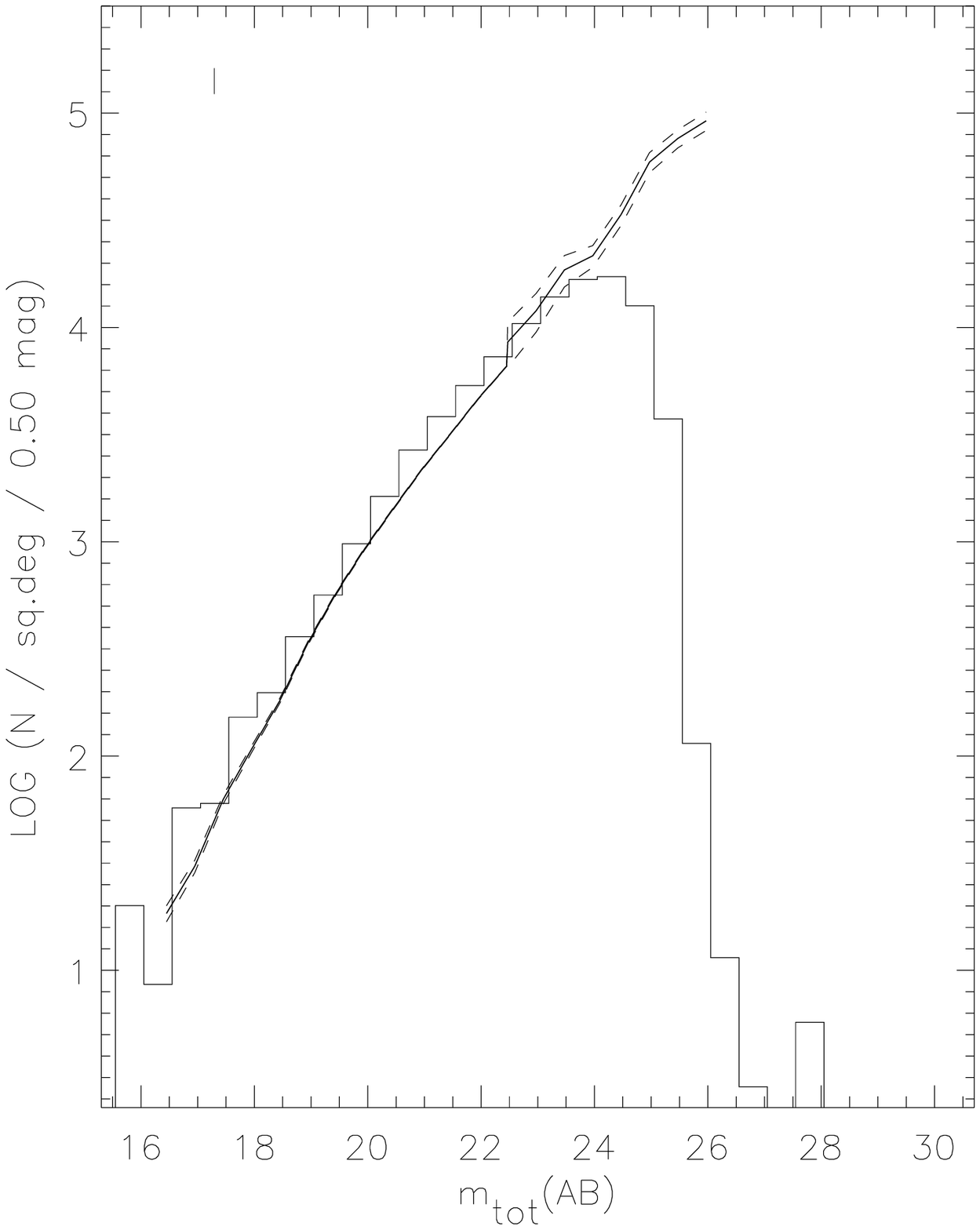}}
\caption{Comparison of galaxy number counts from the present work
(histograms) with those from the following authors: $U-$band:
Guhathakurta \etal (1990); $BVR$-bands: Arnouts \etal (1997) and
Arnouts \etal (1999); $I-$band: Postman \etal (1998) and Arnouts \etal
(1999). The top left panel refers to $U^\prime$, while the top right
panel to $U$. Solid lines represent the mean number counts and the
dashed lines the associated errors.}
\label{fig:countsu}
\end{figure*}

Similarly, Figure~\ref{fig:countsu} shows the comparison of galaxy
counts as a function of the total magnitude, for each of the available
passbands, with those obtained for other data sets. In this case
galaxies were defined as objects with a stellarity index less than
0.95 (0.90 in the $I-$band) or fainter than the classification limit
of each passband. The sample includes only objects with $S/N>3$ and
with the appropriate SExtractor and EIS flags and the normalization of
the counts takes into account the effective area of each catalog. The
$U-$band counts of Guhathakurta \etal (1990) and $I$-band counts of
Postman \etal (1998) have been corrected assuming an $AB$ correction
of 0.8 and 0.45, respectively.  As can be seen the observed counts are
in good agreement with previously published results indicating that
the catalogs being distributed are statistically consistent.

\subsection{Survey Performance}

The characteristics of the data obtained by the present survey are
summarized in Table~\ref{tab:quality} which lists: in column (1) the
filter; in column (2) the seeing of the final co-added image; in
columns (3) and (4) the $5\sigma$ and $3\sigma$ limiting $AB$
magnitude measured within an aperture $2\times$FWHM; and in column (5)
the number of objects with $S/N\geq3$, \ie the number of entries in
the catalogs being distributed. At the $5\sigma$ limiting magnitudes
the fraction of spurious objects with $S/N\geq5$ is estimated to be
$\lsim2\%$ for all bands. The fraction of spurious objects was
estimated by creating catalogs from the survey images multiplied by
$-1$. The mock images were then used as input to the catalog
production and from the comparison between the catalogs of the mock
and real objects the fraction of false-positive detections was
estimated.

\begin{figure}
\centerline{\hbox{\psfig{figure=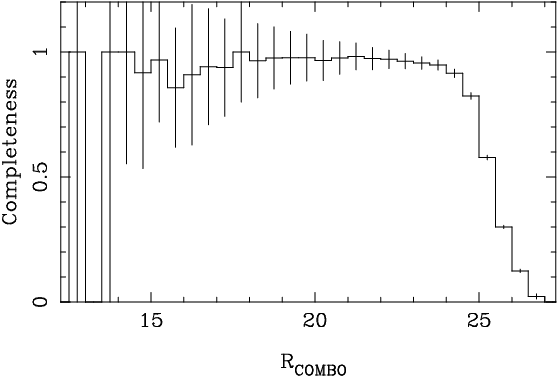,angle=0,width=\columnwidth,clip=}}}
\caption{Estimated completeness, and associated Poisson error, of the
CDF-S $R$-band catalog as a function of the $R_{COMBO}$ magnitude in
the Vega system.}
\label{fig:complete} \end{figure}

The fact that the COMBO data reaches fainter magnitudes can also be
used to obtain a first estimate of the completeness limit of the CDF-S
data in the $R-$band. Figure~\ref{fig:complete} shows the ratio as a
function of the $R_{COMBO}$ of objects in both catalogs. From this
plot one finds that the $R-$band catalog is complete to
$R_{AB}\sim24.7$ and 50\% complete at $R_{AB} \sim 25.7$.  This
estimate is probably conservative. A first indication is that there
are a large number of objects detected only in the data presented
here, most of them at the faint end $R\gsim25.2$. Second, the number
of spurious detections at $R\sim26$ is still relatively small
($\lsim30\%$). Clearly, a determination of the completeness of the
catalog from the CDF-S data themselves is highly desirable and will be
implemented in the near future.

\begin{table}
\caption{Properties of the extracted CDF-S catalogs (AB system). }
\center
\begin{tabular}{ccccc}
\hline\hline
Filter & Seeing & $ m_{lim}(5\sigma)$ 
 & $ m_{lim} (3\sigma)$ &  $N_{obj}$ \\ 
 &  (arcsec) & (mag) & (mag) &   \\
\hline
$U$    &  0.98 & 25.7 & 26.3  & 37094 \\ 
$U'$   & 1.06 & 26.0 & 26.6 &42087 \\
$B$    & 1.02 & 26.4 &  27.0 & 74827 \\ 
$V$    & 0.97 & 25.4 &  26.0 & 45912 \\ 
$R$    &  0.86 & 25.5 &  26.1 & 65148 \\ 
$I$    & 0.93 & 24.7 &  25.3 &43253 \\ 
\hline\hline
\label{tab:quality}
\end{tabular}
\end{table}

\void{
\begin{table*}
\caption{Data summary (magnitudes in the AB system). }
\center
\begin{tabular}{cccccccc}
\hline\hline
Filter & $T_{eff}$ & $N_f$ & Seeing & $ m_{lim}(5\sigma)$ 
 & $ m_{lim} (3\sigma)$ &   $f_{sp}$ & $N_{obj}$ \\ 
 & (sec) &  &  (arcsec) & (mag) & (mag) & &  \\
\hline
$U$    & 47600 & 55 & 0.98 & 25.7 & 26.3 & 0.10  & 37094 \\ 
$U'$   & 43600 & 49 & 1.06 & 26.0 & 26.6 & 0.45 & 42087 \\
$B$    & 24500 & 30 & 1.02 & 26.4 &  27.0 & 0.20 & 74827 \\ 
$V$    &  9250 & 33 & 0.97 & 25.4 &  26.0 & 0.20 & 45912 \\ 
$R$    &  9250 & 33 & 0.86 & 25.5 &  26.1 & 0.30 & 65148 \\ 
$I$    & 26900 & 46 & 0.93 & 24.7 &  25.3 & 0.30 & 43253 \\ 
\hline\hline
\label{tab:quality}
\end{tabular}
\end{table*}
}

The limiting magnitudes reached by the present observations are, in
general, brighter than those originally proposed ($U_{AB}=26.8;
B_{AB}= 26.0; V_{AB}= 26.0; R_{AB}=26.3; I_{AB}=26.0$) For $BVR$ an
increase of the integration time by a factor of two is required to
reach the originally stated $5\sigma$ limits, representing a 50\%
increase in the total integration time requested per pointing.
Certainly more serious is the requirements in $U$ and $I$, even if
more efficient filters are used. From the present data one finds that
the new $U'$ filter has an efficiency at least 35\% larger than the
original $U$.  A similar increase in efficiency in $I$ is possible
using a filter similar to the one available for EMMI for the EIS-WIDE
survey. However, even with the new filters the time required to reach
the desire magnitudes in $U$ and $I$ involve a large increase (a
factor $\gsim 4$) in their already long integration times. Such an
increase would make the desired area coverage unattainable.  On the
other hand, the limiting magnitudes obtained by the present  survey
compare favorably with those reached by EIS-DEEP from which
Lyman-break galaxy candidates drawn from those samples (da Costa \etal
1998; Rengelink \etal 1998) have been successfully confirmed
spectroscopically at $z\gsim2.8$ (Cristiani \etal 2000) strongly
suggesting that the primary goal of the present survey is indeed
possible. A more detailed discussion of this point will be presented
elsewhere when attempts to combine the data obtained using the two
different $U$ filters will be made (Arnouts \etal 2001, Benoist \etal
2001b).

Finally, as mentioned in Section~\ref{sec:observations}, the CDF-S has
already been observed both in optical and infrared passbands as part
of the EIS-DEEP survey using the NTT (Rengelink \etal 1998). These old
observations are being re-analyzed, using the new methods of the
pipeline and compared with the present observations elsewhere
(Benoist \etal 2001a). The infrared data from this survey is
particularly important as it complements the infrared CDF-S survey
recently completed (Vandame \etal 2001).

\section{Summary}
\label{sec:summary}

This paper presents the first optical results of the ongoing
multicolor Deep Public Survey being conducted by the EIS program. The
observations and the data for the first pointing completed as part of
this program which corresponds to the Chandra CDF-S field are
described. The set of products being publicly released includes fully
calibrated pixel maps and source lists covering the CDF-S
region. Given the current interest in this area of the sky both the
optical data presented here, covering $\sim0.25$ square degrees, and
the infrared survey recently released, covering an area of $\sim0.11$
square~degrees, are being made available world-wide. Hopefully, this
will, together with the Chandra X-ray data, contribute to making this
region of the sky a natural target for multi-wavelength observations
in the southern hemisphere and an ideal area for cosmological studies.
Other derived products from the EIS program will be described in
forthcoming papers in this series.

\acknowledgements{ We thank all of those directly or indirectly
involved in the EIS effort.  Our special thanks to K. Meisenheimer and
C. Wolf for making their $U$-band images and $R$-band catalogs
available to EIS, E. Bertin for his quick response to our request for
help, A. Bijaoui for allowing us to use tools developed by him and
collaborators over the years and past EIS team members for building
the foundations of this program.  We would also like to thank
A. Renzini and past and present members of the Working Group for
Public Surveys.  The Guide Star Catalogue-II is produced by the Space
Telescope Science Institute in collaboration with the Osservatorio
Astronomico di Torino.  Space Telescope Science Institute is operated
by the Association of Universities for Research in Astronomy, for the
National Aeronautics and Space Administration under contract
NAS5-26555.  Additional support is provided by the Association of
Universities for Research in Astronomy, the Italian Council for
Research in Astronomy, European Southern Observatory, Space Telescope
European Coordinating Facility, the International GEMINI project and
the European Space Agency Astrophysics Division.   This
publication makes use of data products from the Two Micron All Sky
Survey, which is a joint project of the University of Massachusetts
and the Infrared Processing and Analysis Center/California Institute
of Technology, funded by the National Aeronautics and Space
Administration and the National Science Foundation.  This research has
made use of the {\sc simbad} database, operated at CDS, Strasbourg,
France.  }

{}


\begin{thebibliography}{}



\bibitem[]{arnouts:1997} Arnouts S., de Lapparent V., Mathez
G., \etal, 1997, A\&AS. 124, 163

\bibitem[]{arnouts:1999} Arnouts S., d'Odorico S., Cristiani
 S., \etal, 1999, A\&A, 341, 641

\bibitem[]{arnouts:01c} Arnouts S., \etal, 2001, In preparation

\bibitem[]{benoist:01a} Benoist C., \etal, 2001a, In preparation

\bibitem[]{benoist:01b} Benoist C., \etal, 2001b, In preparation

\bibitem[]{berint:1996} Bertin E., Arnouts S., 1996, A\&AS, 117, 393

\bibitem[]{bertin:97} Bertin, E., 1998, SExtractor, User's guide, v2.0

\bibitem[]{} Bertin E., 2001, {\it private communication}



\bibitem[]{} Cristiani S., Appenzeller I., Arnouts S., \etal,
2000, A\&A 359, 489

\bibitem[]{cutri:00} Cutri R.M., Skrutskie M.F., Van Dyk S.,
  	\etal, 2000, Explanatory Supplement to the 2MASS Second
Incremental Data Release


\bibitem[]{dacosta:98} da Costa L., \etal, 1998, astro-ph/9812105, submitted to A\&A

\bibitem[]{} da Costa L.N.,  et al., 1999, The Messenger 98,36 

\bibitem[]{} da Costa L.N., 2001, in: ``Mining the Sky'', Proc. of
ESO Workshop, Springer-Verlag, In press, astro-ph/0102132


\bibitem[]{djamdji:93} Djamdji  J.P., Bijaoui A., Mani\`{e}re  R., 1993, 
        Photogrammetric Engineering and Remote Sensing, 59, 645

\bibitem[]{giaconni:00} Giacconi R, Rosati P., Tozzi P., \etal, 2001,
 ApJ, 551, 624

\bibitem[]{} Girardi L., 2001, {\it private communication}

\bibitem[]{guha:90} Guhathakurta P., Tyson J.A., Majewski S.R., 1990,
in: Evolution of the universe of galaxies, Astronomical Society of the
Pacific, p. 304

\bibitem[]{}Hatziminaoglou E., \etal, 2001, {\it submitted to A\&A}

\bibitem[]{hoeg:97} H{\o}g E., B\"{a}ssgen G., Bastian U., \etal, 1997,
A\&A, 323, L57

\bibitem[]{} Landolt A.U., 1992,AJ 104,340

\bibitem[]{mclean:00} McLean B.,  Greene G., Lattanzi M., \etal,
2001, in ``Mining the Sky'', Proc. of ESO Workshop, Springer-Verlag, In
press 


\bibitem[]{mendez:96} Mendez R.A., van Altena W.F., 1996, MNRAS 279, 1357


\bibitem[]{perryman:97} Perryman M. A. C., Lindegren L., Kovalevsky J., \etal,
1997, A\&A, 323, L49

\bibitem[]{} Postman M., Lauer T.R., Szapudi I., Oegerle W., 1998, ApJ 506, 33

\bibitem[]{rengelink:98} Rengelink R., \etal, 1998, astro-ph/9812190, submitted to A\&A


\bibitem[]{rosati:01} Rosati P., 2001, in preparation



\bibitem[]{schlegel:98} Schlegel D., Finkbeiner D.,Davis,M., 1998, ApJ, 500,525

\bibitem[]{urban:98} Urban S.E., Corbin T.E., Wycoff G.L.,  1998 AJ 115, 2161

\bibitem[]{} Vandame B., Olsen L.F., J{\o}rgensen H.E., \etal,  
2001, astro-ph/0102300, submitted to A\&A

\bibitem[]{vandame:01} Vandame B., 2001,  in ``Mining the Sky'', Proc. of ESO Workshop, Springer-Verlag, In press 


\bibitem[]{} Wolf C., Meisenheimer  K., Dye  S., Kleinheinrich  M.,
Rix  H.-W., Wisotzki  L., 2000, astro-ph/001247, submitted to
A\&A


\end{thebibliography}
\end{document}